\documentclass[aps,pre,showpacs,floats,twocolumn,superscriptaddress,longbibliography]{revtex4-2}
\renewcommand{\selectlanguage}[1]{}
\usepackage{lipsum}
\usepackage{mathrsfs}
\usepackage{bm,amsbsy,amssymb,amsmath}
\usepackage{caption}
\usepackage{subcaption}
\usepackage{graphics,graphicx,dcolumn,fleqn,epic,eepic,float,tabularx}
\usepackage{multirow,rotate,rotating,color}
\usepackage[utf8]{inputenc}
\newcommand{\figref}[1]{Fig.~\ref{fig:#1}}
\newcommand{\eqnref}[1]{Eq.~(\ref{#1})} 
  \definecolor{tuered}{RGB}{214,0,74}
  \definecolor{tueblue}{RGB}{0,102,204}
  
  \newcommand{\revisedtext}[1]{\textcolor{black}{#1}}

\usepackage{tikz,pgfplots}
\usepackage{relsize}
\tikzset{fontscale/.style = {font=\relsize{#1}}}
\usetikzlibrary{calc}
\graphicspath{{img/}}

\begin{document}
\title{Effect of particle and substrate wettability on evaporation--driven assembly of\\ colloidal monolayers}
  \author{Qingguang Xie}
   \email{q.xie@fz-juelich.de}
\affiliation{Helmholtz Institute Erlangen-N\"urnberg for Renewable Energy (IET-2), Forschungszentrum J\"ulich, Cauerstra{\ss}e 1, 91058 Erlangen, Germany}

\author{Tian Du}
\affiliation{Helmholtz Institute Erlangen-N\"urnberg for Renewable Energy (IET-2), Forschungszentrum J\"ulich, Immerwahrstra{\ss}e 2, 91058 Erlangen, Germany}
\affiliation{Institute of Materials for Electronics and Energy Technology (i-MEET), Department of Materials Science and Engineering, Friedrich-Alexander-Universit\"at Erlangen-N\"urnberg, Martensstraße 7, 91058 Erlangen, Germany}

\author{Christoph J. Brabec}

\affiliation{Institute of Materials for Electronics and Energy Technology (i-MEET), Department of Materials Science and Engineering, Friedrich-Alexander-Universit\"at Erlangen-N\"urnberg, Martensstraße 7, 91058 Erlangen, Germany}
\affiliation{Helmholtz Institute Erlangen-N\"urnberg for Renewable Energy (IET-2), Forschungszentrum J\"ulich, Immerwahrstra{\ss}e 2, 91058 Erlangen, Germany}

\author{Jens Harting}
\email{j.harting@fz-juelich.de}
\affiliation{Helmholtz Institute Erlangen-N\"urnberg for Renewable Energy (IET-2), Forschungszentrum J\"ulich, Cauerstra{\ss}e 1, 91058 Erlangen, Germany}
\affiliation{Department of Chemical and Biological Engineering and Department of Physics, Friedrich-Alexander-Universit\"at Erlangen-N\"urnberg, Cauerstra{\ss}e 1, 91058 Erlangen, Germany}


\begin{abstract}
Assembled monolayers of colloidal particles are crucial for various applications, including optoelectronics, surface engineering, as well as light harvesting, and catalysis. A common approach for
self-assembly is the drying of a colloidal suspension film on a solid substrate using technologies such
as printing and coating. However, this approach often presents challenges such as low surface coverage, stacking faults, and the formation of multiple layers. We numerically investigate the influence
of substrate and particle wettability on the deposited pattern. Higher substrate wettability results
in a monolayer with a hexagonal arrangement of deposited particles on the substrate. Conversely,
lower substrate wettability leads to droplet formation after the film ruptures, leading to the formation of particle clusters. Furthermore, we reveal that higher particle wettability can mitigate the
impact of the substrate wettability and facilitate the formation of highly ordered monolayers. We
propose theoretical models predicting the surface coverage fraction dependent on particle volume
fraction, initial film thickness, particle radius, as well as substrate and particle wettability, and
validate these models with simulations. Our findings provide valuable insights for optimizing the
deposition process in the creation of assembled monolayers of colloidal particles.
\end{abstract}

\maketitle

\section{Introduction}
The assembly of monolayers of colloidal particles is relevant 
in various scientific and technological domains, namely catalysis, photovoltaics, sensors, nanomedicine and batteries~\cite{Xia2000693,zhang_self-assembly_2009,lotito_approaches_2017,ye_two-dimensionally_2011,ocwieja_formation_2017,betancourt_micro-_2006}. Self-assembled monolayers are used as a mask to fabricate ordered nanostructures in colloidal lithography~\cite{sinitskii_patterning_2010}. Moreover, self-assembled monolayers can significantly influence the properties of the overall structure. For example, solution-processed thin films based on metal oxide nanoparticles were widely employed in organic or hybrid optoelectronic devices, as they are the particularly versatile materials used as interfacial buffer layers to resolve energetic misalignment in organic electronics~\cite{shahid2022,chavali2019}. Their applications range from charge injection layers in light-emitting diodes, gate layers in organic field-effect transistors, charge extraction layers for organic solar cells~\cite{greiner_thin-film_2013} to more recently charge-transporting layers in perovskite solar cells~\cite{shin_metal_2019}. 
Additionally, broadband light absorption enhancement has been observed in ultrathin film crystalline silicon solar cells with the incorporation of polystyrene colloidal monolayers~\cite{wang_broadband_2016}.

The assembly of monolayers of colloidal particles is usually done by drying a suspension film on a substrate~\cite{denkov_mechanism_1992,reculusa_synthesis_2003,kaliyaraj_selva_kumar_mini-review_2020}.
Here, one utilizes techniques such as printing and coating, which are easy-to-use, low-cost and scalable~\cite{dimitrov_continuous_1996,steinberger_challenges_2024}. Furthermore, the flat fluid-fluid interface prevents capillary flow and radial movement of particles, usually encountered in 
drying a colloidal suspension droplet due to contact line pinning~\cite{deegan_capillary_1997, mayarani_colloidal_2021}. However, achieving a uniform deposition pattern from drying a thin film of colloidal particles also poses formidable challenges due to several inherent complexities. The deposition of particles is susceptible to interparticle forces such as van der Waals attraction and capillary forces, as well as particle-fluid and particle-substrate interactions~\cite{lotito_approaches_2017,van_dommelen_surface_2018,roach_controlling_2022}. 
Zargartalebi et al.~\cite{zargartalebi_self-assembly_2022} produced highly ordered particle deposits by drying a suspension film on a superhydrophilic substrate surrounded by a neutrally wetting mold with low roughness. They claimed that a meniscus-free interface and a hydrophilic substrate are required to produce highly ordered particle assemblies.
Fujita et al.~\cite{fujita_direct_2015} numerically addressed the effect of particle wettability on the deposition process on a hydrophilic substrate. 
Similarly, Mino et al.~\cite{mino_numerical_2022} simulated the drying process of a colloidal suspension on a wetting substrate. Their findings revealed that particles with higher wettability exhibited slower aggregation. However, the effect of substrate wettability and its interplay with particle wettability on the deposited pattern are neglected, despite their pivotal roles in determining the process of particle deposition.

In this paper, we perform simulations of drying a colloidal suspension film utilizing a coupled lattice Boltzmann and discrete element method. The lattice Boltzmann method is a powerful tool to model fluid flow involving solvent evaporation~\cite{HXH17}. The particles are discretized on the lattice and are coupled with a fluid solver through a momentum exchange approach~\cite{ladd-verberg2001,Jansen2011}.
Initially, we compare the temporal evolution of the evaporated mass during the drying process of both a pure liquid film and a colloidal suspension film on a substrate 
with its respective analytical prediction. Subsequently, we explore the particle deposition resulting from the drying of a colloidal suspension film, manipulating the substrate wettability. 
On a well wetting substrate, 
the film undergoes drying and dewetting, resulting in the formation of a monolayer deposit during the evaporation process. Conversely, 
lower substrate wettability leads to film rupture and droplet formation, leaving behind particle clusters after drying. 
Importantly,
our findings furthermore demonstrate that the particle wettability 
has the capability to mitigate the influence of the substrate wettability. 
We propose theoretical models to predict the surface coverage fraction, considering the particle volume fraction and incorporating the wetting properties of both particle and substrate. 
These models are in good agreement with our simulation results.

\section{Methods}
\label{sec:method}
We employ the lattice Boltzmann method (LBM), a computational technique used for modeling fluid dynamics at the mesoscopic scale, offering a unique and versatile approach to simulate complex fluid flow phenomena~\cite{bib:succi-01}. Unlike traditional methods based on solving the Navier-Stokes equations directly, the LBM is rooted in kinetic theory, employing a lattice to represent fluid particles and their collisions. 
In the regime of small Knudsen and Mach numbers, the Navier-Stokes equations are reinstated~\cite{bib:succi-01}. 
Over the last two decades, the LBM has proven itself as a robust tool for numerically simulating fluid flows~\cite{bib:succi-01}. It has been expanded to model multiphase/multicomponent fluids~\cite{Shan1993, Liu2016} and suspensions of particles with varying shape and wettability~\cite{ladd-verberg2001, Jansen2011, Gunther2013a, Xie2015}. The inherent parallelizability and adaptability of the LBM to irregular geometries make it particularly advantageous for studying intricate fluid dynamics scenarios. 
In the subsequent discussion, we outline relevant details and direct readers to the relevant literature for an in-depth description of the method and our implementation~\cite{Jansen2011, Frijters2012, HXH17, Liu2016}.

We utilize the pseudopotential multicomponent LBM of Shan and Chen~\cite{Shan1993} 
with a D3Q19 lattice~\cite{Qian1992}.
Here, two fluid components are modelled 
by following the evolution of each distribution function 
discretized in space and time according to the lattice Boltzmann equation,
\begin{eqnarray}
  \label{eq:LBG}
  f_i^c(\mathbf{x} + \mathbf{e}_i \Delta t , t + \Delta t)&= &f_i^c(\mathbf{x},t) - \frac{\Delta t} {\tau^c} [  f_i^c(\mathbf{x},t) - \nonumber \\
  &&f_i^\mathrm{eq}(\rho^c(\mathbf{x},t),\mathbf{u}^c(\mathbf{x},t))]
  \mbox{\,,}
\end{eqnarray}
where $i=0,...,18$. $f_i^c(\mathbf{x},t)$ are the single-particle distribution
functions for fluid component $c=1$ or $2$, and $\mathbf{e}_i$ is the discrete
velocity in the $i$th direction.  $\tau^c$ is the relaxation time for component
$c$ and determines the viscosity.  The macroscopic densities and velocities for
each component are defined as $\rho^c(\mathbf{x},t) = \rho_0
\sum_if^c_i(\mathbf{x},t)$, where $\rho_0$ is a reference density, and
$\mathbf{u}^c(\mathbf{x},t) = \sum_i  f^c_i(\mathbf{x},t)
\mathbf{e}_i/\rho^c(\mathbf{x},t)$, respectively. 
Here, $f_i^\mathrm{eq}$ is the second-order equilibrium distribution function defined as
\begin{eqnarray}
  \label{eq:eqdis}
  f_i^{\mathrm{eq}}(\rho^c,\mathbf{u}^c) &=& \omega_i \rho^c \bigg[ 1 + \frac{\mathbf{e}_i \cdot \mathbf{u}^c}{c_s^2} \nonumber \\
  && - \frac{ \left( \mathbf{u}^c \cdot \mathbf{u}^c \right) }{2 c_s^2} + 
  \frac{ \left( \mathbf{e}_i \cdot \mathbf{u}^c \right)^2}{2 c_s^4}  \bigg]
  \mbox{\,,}
\end{eqnarray}
where $\omega_i$ is a coefficient depending on the direction: $\omega_0=1/3$
for the zero velocity, $\omega_{1,\dots,6}=1/18$ for the six nearest neighbors
and $\omega_{7,\dots,18}=1/36$ for the nearest neighbors in diagonal direction.
$c_s = \frac{1}{\sqrt{3}} \frac{\Delta x}{\Delta t}$ is the speed of sound.

For convenience, we choose the lattice constant $\Delta x$, the timestep $
\Delta t$, the reference density $\rho_0 $ and the relaxation time $\tau^c$ to be
unity, which leads to a kinematic viscosity $\nu^c$ $=$ $\frac{1}{6}$ in
lattice units.

The pseudopotential multicomponent model introduces a mean-field interaction force 
\begin{equation}
  \label{eq:sc}
\!\!\!\!\!  \mathbf{F}^c(\mathbf{x},t) = -\Psi^c(\mathbf{x},t) \sum_{\bar{c}} \sum_{i} \omega_i g_{c\bar{c}} \Psi^{\bar{c}}(\mathbf{x}+\mathbf{e}_i,t) \mathbf{e}_i
\end{equation}
between fluid components $c$ and $\bar{c}$~\cite{Shan1993}, 
in which $g_{c\bar{c}}$ is a coupling constant, eventually leading to a demixing of the fluids.
We denote $\gamma$ as the surface tension of the interface.
$\Psi^c(\mathbf{x},t)$ is an ``effective mass'', chosen as the functional form
\begin{equation}
  \label{eq:psifunc}
  \Psi^c(\mathbf{x},t) \equiv \Psi(\rho^c(\mathbf{x},t) ) = 1 - e^{-\rho^c(\mathbf{x},t)}
   \mbox{\,.}
\end{equation}
This force $\mathbf{F}^c(\mathbf{x},t)$ is then applied to the component $c$ by adding a 
shift $\Delta \mathbf{u}^c(\mathbf{x},t) =\frac{\tau^c \mathbf{F}^c(\mathbf{x},t)}{\rho^c(\mathbf{x},t)}$ 
to the velocity $\mathbf{u}^c(\mathbf{x},t)$ in the equilibrium distribution.

When the interaction parameter $g_{c\bar{c}}$ in~\eqnref{eq:sc} is appropriately selected, the separation of components occurs, leading to the formation of distinct phases. Each component segregates into a denser majority phase with a density of $\rho_{ma}$ and a lighter minority phase with a density of $\rho_{mi}$. The diffusive nature of the interface prevents the occurrence of stress singularities at the moving contact line, a phenomenon typically observed in sharp-interface models.

To model substrate wettability, we introduce an interaction force between the fluid and wall, inspired by the work of Huang et al.~\cite{Huang2007}, as
\begin{equation}
\mathbf{F}^c(\mathbf{x}) =  - g^{wc} \Psi^c (\mathbf{x}) \sum_{i} \omega_{i} s(\mathbf{x} + \mathbf{e}_{i}) \mathbf{e}_{i}
\mbox{,}
\label{eq:rhowall2}
\end{equation}
where $g^{wc}$ is a constant. Here, $s(\mathbf{x} + \mathbf{e}_{i})= 1$ if
$\mathbf{x} + \mathbf{e}_{i}$ is a solid lattice site, and $s(\mathbf{x} +
\mathbf{e}_{i})=0$ otherwise. 

To induce evaporation, we enforce a constant value $\rho^c_H$ for the density of component $c$ at the boundary sites $\mathbf{z}_H$ by specifying the distribution function of component $c$ as~\cite{HXH17}
\begin{equation}
f_i^c(\mathbf{z}_H,t) = f_i^\mathrm{eq}\left(\rho_H^c,\mathbf{u}^c_H(\mathbf{z}_H,t)\right).
\end{equation}
Here, $\mathbf{u}^c_H(\mathbf{z}_H,t)=0$. If the prescribed density $\rho_H^c$ is lower than the equilibrium minority density $\rho_{mi}^c$, a density gradient is established in the vapor phase of component $c$. This gradient prompts the diffusion of component $c$ toward the evaporation boundary. 
The diffusion coefficient of component $c$ is given as 
 $D_c = c^{2}_{s}(\tau-\frac{1}{2}) \frac{\rho_{\bar{c}}}{\rho_c+\rho_{\bar{c}}} - \frac{c_{s}^{2}\rho_c g_{\bar{c}c}\Psi'_c\Psi_{\bar{c}}}{\rho_c+\rho_{\bar{c}}}$, 
where $\Psi'=d\Psi/d\rho$~\cite{shan_multicomponent_1995,HXH17}. It is important to note that our evaporation model is diffusion-dominated, which is validated in our prior work~\cite{HXH17}.

The colloidal particles are discretized on the fluid lattice, and their interaction with the fluid species is established through a modified bounce-back boundary condition, a method pioneered by Ladd and Aidun~\cite{ladd-verberg2001, AIDUN1998}. The motion of the particles is governed by classical equations of motion:
\begin{eqnarray}
\mathbf{F}{\mathrm{p}}=m \frac{d\mathbf{u}_{\mathrm{p}}}{dt} \mbox{\,} 
\label{eq: newton}
\end{eqnarray}
Here, $\mathbf{F}{\mathrm{p}}$ represents the total force acting on a particle with mass $m$, and $\mathbf{u}_{\mathrm{p}}$ is the particle's velocity. The trajectory of a colloidal particle is updated using a leap-frog integrator. \revisedtext{ Given that we treat particles as rigid spheres, we neglect rotational motion and particle deformation.} 

We introduce a ``virtual'' fluid within the outer shell of the particle, with an amount $\Delta \rho_p$~\cite{Jansen2011, Frijters2012}, expressed as
\begin{eqnarray}
 \rho_{\mathrm{virt}}^1(\mathbf{x},t)  &=& \overline{\rho}^1(\mathbf{x},t) + \Delta\rho_p \mbox{\,, } \\
 \rho_{\mathrm{virt}}^2(\mathbf{x},t) & =& \overline{\rho}^2(\mathbf{x},t) - \Delta\rho_p  \mbox{\,. }
 \label{eq:md_colour}
\end{eqnarray}
$\overline{\rho}^1(\mathbf{x},t)$ and $\overline{\rho}^2(\mathbf{x},t)$ represent the averages of the density of neighboring fluid nodes for components 1 and 2, respectively. 
\revisedtext{
The virtual fluid inside the particles is incorporated into the calculation of the Shan-Chen interaction force~\eqnref{eq:sc}, which ensures a proper force balance and prevents the formation of an artificial fluid density layer around the particles~\cite{Jansen2011}. The Shan-Chen interaction between the particles and the surrounding fluids can be tuned by adjusting the density of the local virtual fluid. Increasing the density of one component by an amount $\Delta\rho_p$ makes the particle surface “prefer” that fluid over the other.
The parameter $\Delta\rho_p$, referred to as the “particle colour”, governs the particle’s wettability and thus determines its contact angle. The contact angle varies approximately linearly with the particle colour, with the slope of this relationship depending on the particular simulation parameters used~\cite{Jansen2011}.
} A particle color of $\Delta\rho_p = 0$ corresponds to a contact angle of $\theta = 90^{\circ}$, indicating a neutrally wetting particle.

The exchange of momentum between particles and the fluid accounts for hydrodynamic forces, including drag and lift forces. Our model accurately captures lubrication interactions when the distance between two particles is at least one lattice site. However, when the separation is less than one lattice site, a lubrication correction is applied~\cite{ladd-verberg2001, JHT16, KHV10}:
\begin{equation}
 \mathbf{F}_{ij} = -\frac{3\pi \mu R^2}{2} \hat{\mathbf{r}}_{ij} \hat{\mathbf{r}}_{ij} \cdot (\mathbf{u}_i-\mathbf{u}_j) \left( \frac{1}{r_{ij}-2R}-\frac{1}{\Delta_c} \right)\,,
 \end{equation}
Here, $R$ represents the radius of the particle, $\mathbf{\hat{r}}_{ij}=\frac{\mathbf{r}_{i}-\mathbf{r}_{j}}{|\mathbf{r}_{i}-\mathbf{r}_{j}|}$ is a unit vector pointing from the center of one particle to the center of the other, and $r_{ij}$ is the distance between particles $i$ and $j$. The velocities of particles are denoted by $\mathbf{u}_i$ and $\mathbf{u}_j$. The constant $\Delta_c$ is chosen as $\Delta_c=2/3$.

The Van der Waals forces acting between two spherical particles with identical radii \revisedtext{$R$} 
are modelled as
\begin{equation}
    F_{vdW}=\frac{A_H R}{12 (r_{ij}-2R)^{2}}, \quad\mbox{for} \quad r_s \le  r_{ij}\le r_c\,,
    \label{eq:vdw}
\end{equation}
where $A_H$ is the Hamaker constant, and $r_s$ and $r_c$ are the cut off radii. 
We set $r_s=2R+0.2$ and $r_c = 2R+1$ in our simulations.

To prevent the overlap of particles, we introduce a Hertz potential~\cite{hertz1881}: 
\begin{equation}
  \label{eq:hertz-potential}
  \phi_H = \left\{\begin{matrix}
K_H(2R-r_{ij})^{\frac{5}{2}}\quad\mbox{for}\quad r_{ij}\le 2R
\\ 
0, \quad\mbox{otherwise} 
\end{matrix}\right.
\end{equation}
Here, $K_H$ is the force constant and is chosen to be $K_H = 100$. \revisedtext{The Hertz hard-sphere potential governs particle-particle interactions at close contact, eliminating the need for an explicit contact model.}

For the interactions between particles and a substrate, 
the lubrication forces between particles and walls are modeled similarly to the lubrication forces between particles themselves.
Additionally, \revisedtext{to prevent particle-substrate penetration and model particle adsorption onto the substrate,} we implement the
Lennard-Jones (LJ) potential between particles and a substrate as
\begin{equation}
\!\!\!\!\!\!\!\!\! \phi_{LJ} = 4\epsilon \left[  \left(\frac{\sigma}{r_{iw}}\right)^{12} - \left(\frac{\sigma}{r_{iw}}\right)^6 \right] \quad\mbox{for}\quad r_{iw}\le r_{c2}\,,
\label{eq:lj}
\end{equation}
where $\epsilon$ is the depth of the potential well, $\sigma$ is the finite
distance at which the inter-particle potential is zero, $r_{iw}$ is the
distance between a particle center and the substrate surface and $r_{c2}$ is the cut off radius.  We set $\sigma$
equal to the particle radius $R$ and $r_{c2}=2.5R$ in all simulations.

Our numerical models were validated previously by comparing the capillary forces between neighbouring particles at fluid interfaces~\cite{Xie2015,XDH16}, the evolution of interface position when drying a purely liquid film and a floating droplet~\cite{HXH17}, as well as the velocity field in an evaporating sessile droplet with theoretical analysis and experimental observations~\cite{XH18a}. We note that the evaporation-driven dynamics during the drying of a colloidal film are primarily governed by vapor diffusion through the surrounding fluid phase, whereas the specific properties of the surrounding fluid have a negligible influence on the overall behavior. Therefore, we employed a multicomponent model instead of a multiphase model to ensure numerical stability.
 
The parameter values chosen in our simulations correspond to particles with a radius \revisedtext{on the order of $100$ $nm$} in water, which has a dynamic viscosity of $\eta_w = 10^{-3}$ $ \mathrm{Pa \cdot s}$, a mass density of $\rho_w = 10^3$ $ \mathrm{kg/m^{3}}$, and a surface tension of $\sigma_w = 7.2 \times 10^{-2}$ $ \mathrm{N/m}$. We consider a system in which particle diffusion is much slower than the movement of the liquid interface driven by evaporation, implying a fast-evaporation regime characterized by a high Péclet number, $Pe \gg 1$, defined as the ratio of the characteristic timescales of particle diffusion and interface movement.  Consequently, the Brownian motion of colloidal particles is neglected in our simulations. The droplet shape is dominated by surface tension, corresponding to a small Bond number ($Bo \ll 1$), such that gravitational effects are negligible. Furthermore, we assume the particles have a density similar to that of the liquid, so gravitational forces are not applied to the particles.

\section{Results and discussion}
\label{sec:results}
\begin{figure}[h]
 \centering
    \includegraphics[width=0.4\textwidth]{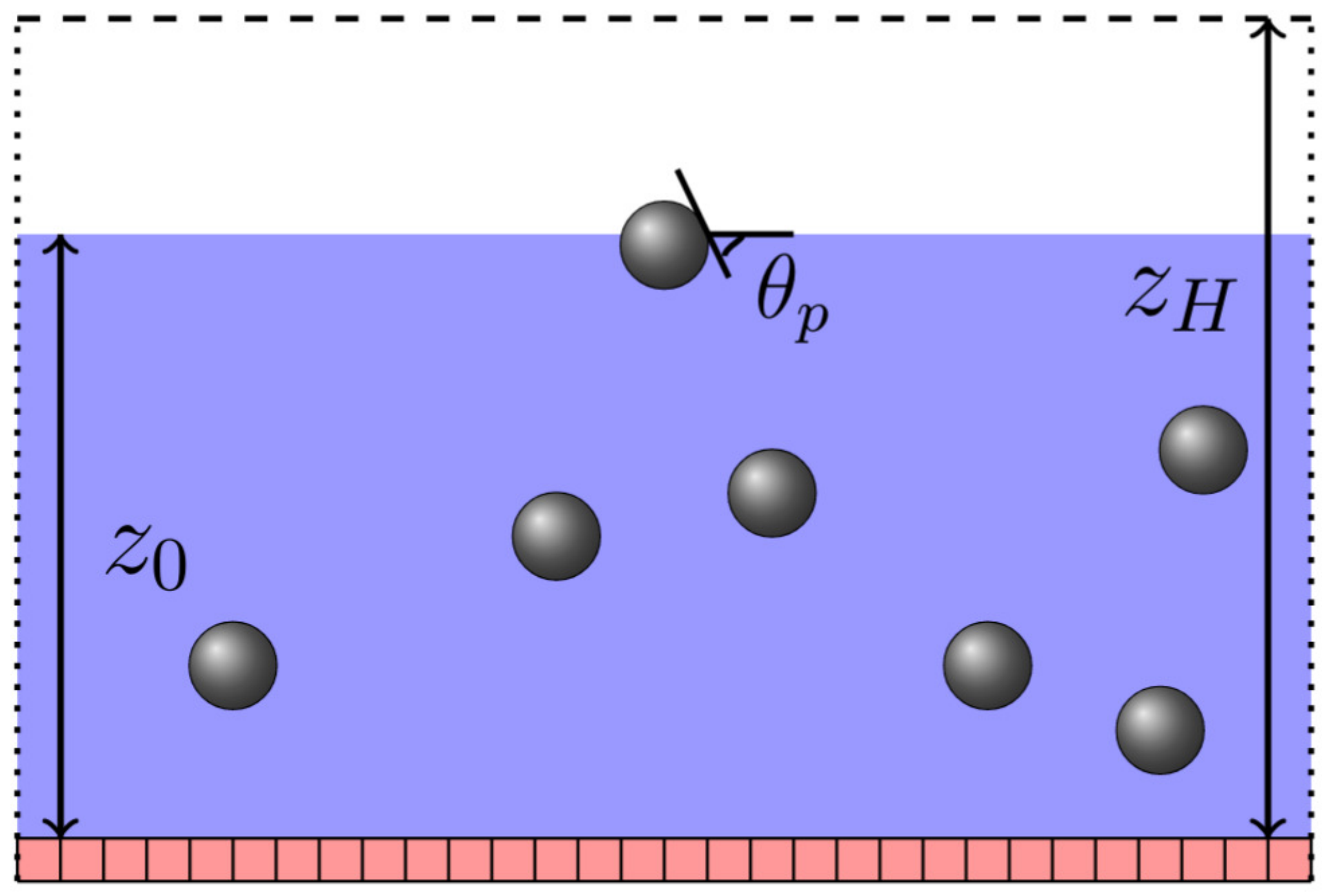}
\caption{Illustration of a thin colloidal suspension film at the initial state. 
The initial height of the film is $z_0$, and the contact angle of the particle is $\theta_p$.
We place a substrate with defined wettability at the bottom, whereas the boundaries normal
to the substrate are periodic (dotted lines). After equilibration, we apply evaporation
boundary conditions at the top of the system (shown by the dashed line). 
The distance between the evaporation boundary and the substrate is $z_H$.
}
\label{fig:geo}
\end{figure}
We investigate the evaporation of a planar film on a solid substrate, as illustrated in~\figref{geo} and 
perform simulations with a system size of $128\times 128 \times 128$ lattice nodes, unless specified otherwise.
One portion of the system is filled with fluid $c$, while the other contains an equally dense fluid $\bar{c}$. 
This setup results in the emergence of a fluid-fluid interface at position $z_0$. 
The position of the interface is defined as the position where $\rho_c = \rho_{\bar{c}}$. 
For the interaction between the fluids, we choose a strength of $g_{c\bar{c}}=3.6$ in~\eqnref{eq:sc}, yielding a diffusive interface with a thickness of $\approx 5$ lattice nodes and a corresponding surface tension $\gamma \approx 0.47$. A wall with a thickness of $2$ lattice nodes is positioned at the bottom, parallel to the interface, and is enforced with simple bounce-back boundary conditions. The boundaries perpendicular to the substrate are set to be periodic. The van der Waals force between particles is applied, as described by~\eqnref{eq:vdw} with $A_H=0.0467$. A Lennard-Jones potential is employed between particles and the substrate, following~\eqnref{eq:lj}. 
We note that the friction force between the particles and the substrate can significantly influence particle deposition. However, in this context, we assume zero friction between the particle and the substrate, given the dominance of capillary forces~\cite{XH18a}.

\subsection{Drying dynamics of a film}

We start with investigating the drying dynamics of both, a pure liquid film and a colloidal suspension film. 
The contact angle of the substrate is fixed at $\theta_s=90^{\circ}$ and after allowing the system to equilibrate without evaporation, we impose the evaporation boundary condition and monitor the evaporated mass over time.

\begin{figure}[h!]
    \centerline{
    \includegraphics[width=0.45\textwidth]{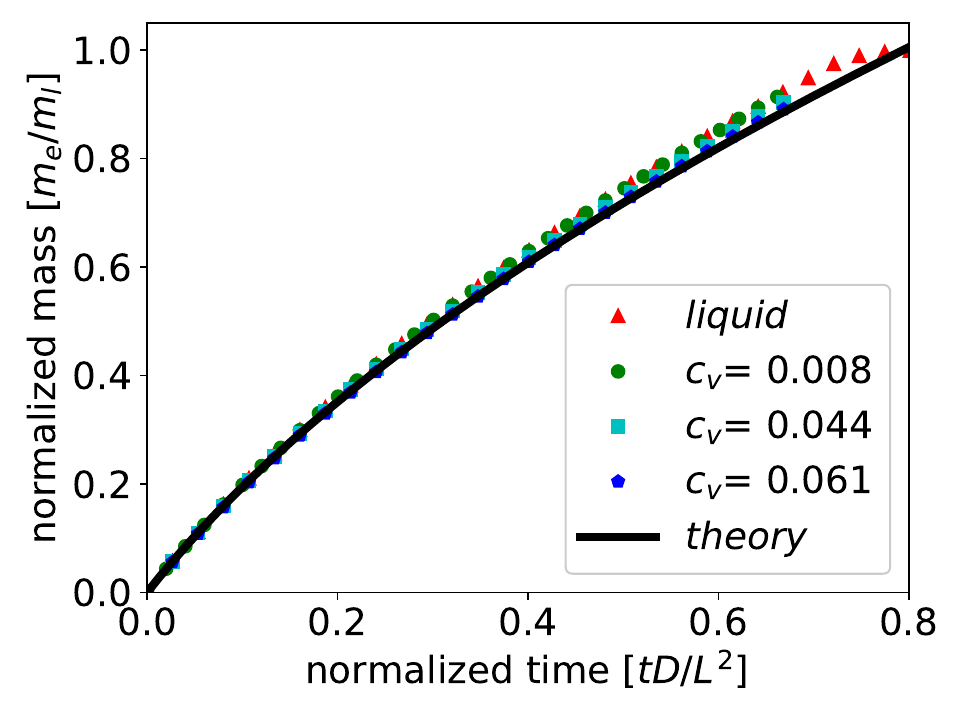}}
    \caption{Time evolution of evaporated mass from a drying colloidal suspension film for particle volume fractions $c_v=0.008$ (circles), $c_v=0.044$ (squares), and $c_v=0.061$ (pentagons), in comparison with evaporating a pure liquid film (triangles) and the theoretical prediction $\eqnref{eq:evap_mass}$ (solid line). 
    The evaporated mass $m_e$ is normalized by the initial total mass $m_l$ and the time is normalized by the diffusivity $D$ of the liquid and the size of the system $L$.
    }
    \label{fig:evap_mass}
\end{figure}

\figref{evap_mass} illustrates the temporal evolution of the evaporated mass for both, a pure liquid film (triangles) and a colloidal suspension film, considering various particle volume fractions $c_v=0.008$ (circles), $c_v=0.044$ (squares), and $c_v=0.061$ (pentagons).
We use particles with a radius of $R=6$ lattice nodes (\revisedtext{corresponding approximately to the order of $100$ $nm$}) to eliminate the effects of the diffusive interface, 
such that the particles effectively cover the interface rather than behaving as if they are immersed within it. 
Initially, the particles are randomly dispersed in the liquid. As drying progresses, an increasing number of particles accumulates at the interface, with the maximum interface coverage fraction ranging from $9.7\%$ to $70\%$, depending on the selected volume fractions.

The evaporated mass is normalized as $m_e /m_l$, where $m_l=\rho_c L^2 z_0$ is the initial total mass of liquid in the case of a purely liquid film. 
The time is normalized with the diffusivity $D \approx 0.117$ of the fluid, and the length $L=128$ of the system.
While the presence of particles at interfaces is expected to influence liquid evaporation, we surprisingly observe overlapping curves across all cases. 
In the following, we present a theoretical analysis to elucidate the observed phenomena.

By assuming the formation of a linear density gradient in the vapor phase, 
the evaporation flux can be estimated as~\cite{HXH17} 
\begin{equation}
\mathbf{j} = -D \nabla \rho = -D \frac{\rho_{H}-\rho_{mi}}{z_H-z_i} \mathbf{n}\,,
\end{equation}
where $\mathbf{n}$ is the normal vector of the interface. 
The time derivative of the mass of the liquid is
\begin{equation}
    \frac{dm}{dt} =  A |\mathbf{j}|\,,
    \label{eq:dmdt1}
\end{equation}
in which $A$ is the area of the interface.
In the case that the thickness of the diffusive interface is significantly smaller than the system size,
the total mass of the system is approximately 
\begin{equation}
    m = A \left[z_i \rho_{ma} +  (z_H-z_i)(\rho_{mi}-\rho_{H})/2 \right]
\end{equation}
and we obtain 
\begin{equation}
\frac{dm}{dt} = A\left[\rho_{ma}  - (\rho_{mi}-\rho_{H})/2\right] \frac{dz_i}{dt}\,.
\label{eq:dmdt2}
\end{equation}
By comparing~\eqnref{eq:dmdt1} and~\eqnref{eq:dmdt2},
\begin{eqnarray}
   A[\rho_{ma}  - (\rho_{mi}-\rho_{H})/2] \frac{dz_i}{dt} = -AD \frac{\rho_{H}-\rho_{mi}}{z_H-z_i}
\end{eqnarray}
we obtain the evolution of the interface position $z_i$ as
\begin{equation}
    \frac{dz_i}{dt} = \frac{D (\rho_{H}-\rho_{mi})}{(z_H-z_i)[\rho_{ma}  - (\rho_{mi}-\rho_{H})/2]}\,.
\end{equation}
The position of the interface $z_i$ follows
\begin{equation}
    z_i = z_H - \left[(z_H-z_0)^2+2\frac{D (\rho_{mi}-\rho_{H}) }{\rho_{ma}  - \frac{\rho_{mi}-\rho_{H}}{2} } t\right]^{1/2}
\end{equation}
and the evaporated mass $m_e$ is
\begin{eqnarray}
    m_e &=& A\rho_{ma} ( z_0-z_i) = A\rho_{ma} ( z_0-z_H) \nonumber \\ 
    &&+ A\rho_{ma} \left[(z_H-z_0)^2+\frac{2D (\rho_{mi}-\rho_{H}) }{\rho_{ma}  - \frac{\rho_{mi}-\rho_{H}}{2}} t\right]^{1/2}\,.
    \label{eq:evap_mass}
\end{eqnarray}
\figref{evap_mass} shows that the analytical prediction~\eqnref{eq:evap_mass} (solid line) agrees well with simulation results (symbols). 
We note that the evaporated mass can be written in an alternative form as 
\begin{equation}
    m_e = \int_0^t A |\mathbf{j}| dt\,.
\end{equation}
Traditionally, the interface area $A$ is considered the effective evaporation area. However, using this approach would imply a slowdown in evaporation with an increasing particle volume fraction, as particles occupy a portion of the interface. However, if the liquid passes the interface much faster than it would through pure diffusion 
and if the particle radius is much smaller than the system size, $R\ll z_H$, the vapor phase just above the particles saturates immediately and the evaporation flux remains constant. Consequently, the effective evaporation area remains constant, even in the presence of particles at the interface. 
\revisedtext{As is commonly encountered in the printing and coating processes of catalyst inks or solutions of functional materials used in organic or perovskite solar cells, $z_H$ can be estimated from the initial wet film thickness, which typically ranges from a few tens to hundreds of micrometers—much larger than the particle size, and well within the scope of our proposed model.
Furthermore,} our findings offer a possible explanation that the theoretical analysis of the velocity field, derived from the drying of a pure liquid droplet, successfully predicts the velocity profile within a drying colloidal suspension droplet~\cite{deegan_contact_2000,popov_evaporative_2005,marin_order--disorder_2011}.

\subsection{Effect of substrate wettability }
\begin{figure}[t]
    \centering
    \includegraphics[width=0.45\textwidth]{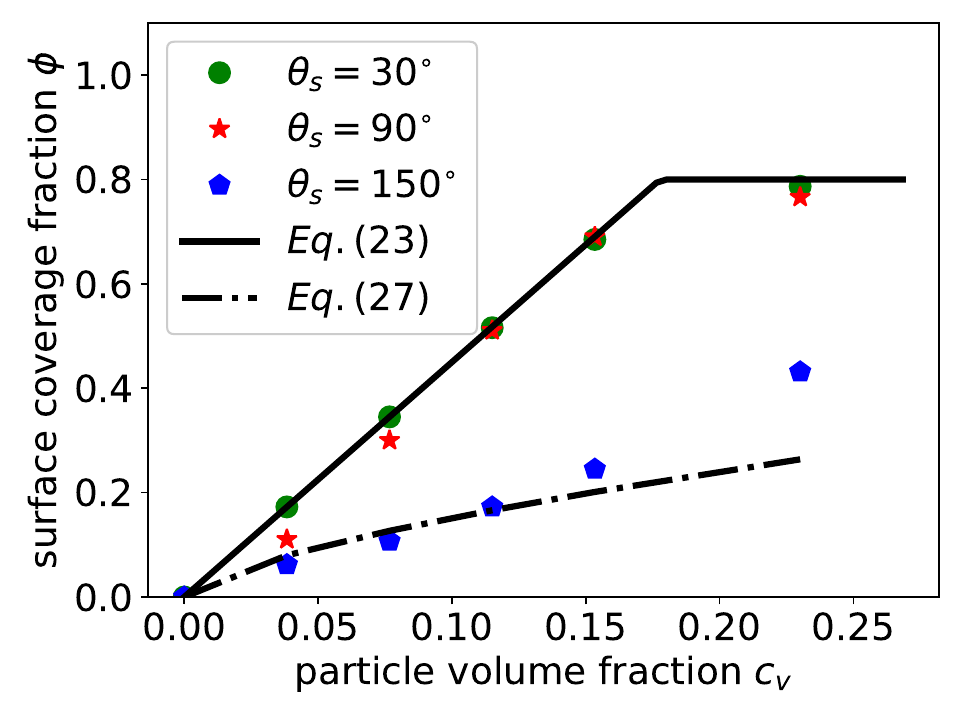}
    \caption{Surface coverage fraction $\phi$ as a function of particle volume fraction $c_v$ for a substrate with different contact angles $\theta_s=30^{\circ}$, $\theta_s=90^{\circ}$, and $\theta_s=150^{\circ}$. The contact angle of particles is fixed to $\theta_p=90^{\circ}$.}
    \label{fig:theta_s}
\end{figure}
In the following, we study the impact of the substrate wettability 
on the deposition process. We characterize the wettability of the substrate by the contact angle of a droplet on the substrate: a lower contact angle indicates higher wettability, while a higher contact angle corresponds to lower wettability.
We initiate the film with a particle volume concentration $c_v$ and choose particles with a radius of 3 to ensure an adequate number of particles while saving computational time. Since our focus is on the effect of substrate contact angle here, the particle radius does not significantly impact the results.
The particles are neutrally wetting (contact angle $\theta_p = 90^{\circ}$) and we vary the substrate contact angle, examining cases with $\theta_s = 30^{\circ}$, $\theta_s=90^{\circ}$ and $\theta_s = 150^{\circ}$. 

In~\figref{theta_s}, we compare the surface coverage fraction $\phi$ as a function of particle volume fraction 
for the different substrate contact angles. \revisedtext{To save computational cost, we limit simulations to systems with particle volume fraction $\phi < 0.25$.}  The surface coverage fraction, $\phi$, is calculated after the solvent has evaporated, based on the number of particles, $N_p^s$, attached to the substrate using the formula $\phi = \frac{\pi N_{p}^{s} R^2}{S}$ ($S$ corresponds to the area of the substrate).
In all cases, the surface coverage fraction increases with increasing particle volume fraction. 
At lower volume fraction $c_v<0.08$, the surface coverage fraction with a substrate contact angle $\theta_s=30^{\circ}$ is slightly higher than that with a contact angle $\theta_s=90^{\circ}$, but the curves overlap at higher volume fraction $c_v>0.1$. Throughout the entire range, 
the surface coverage fraction at contact angles $\theta_s=30^{\circ}$ and $\theta_s=90^{\circ}$ is larger than that with a contact angle $\theta_s=150^{\circ}$. 

\begin{figure*}[ht!]
    \centering
    \captionsetup[subfigure]{justification=centering}
    \begin{subfigure}{.15\textwidth}
		\includegraphics[width=0.95\textwidth]{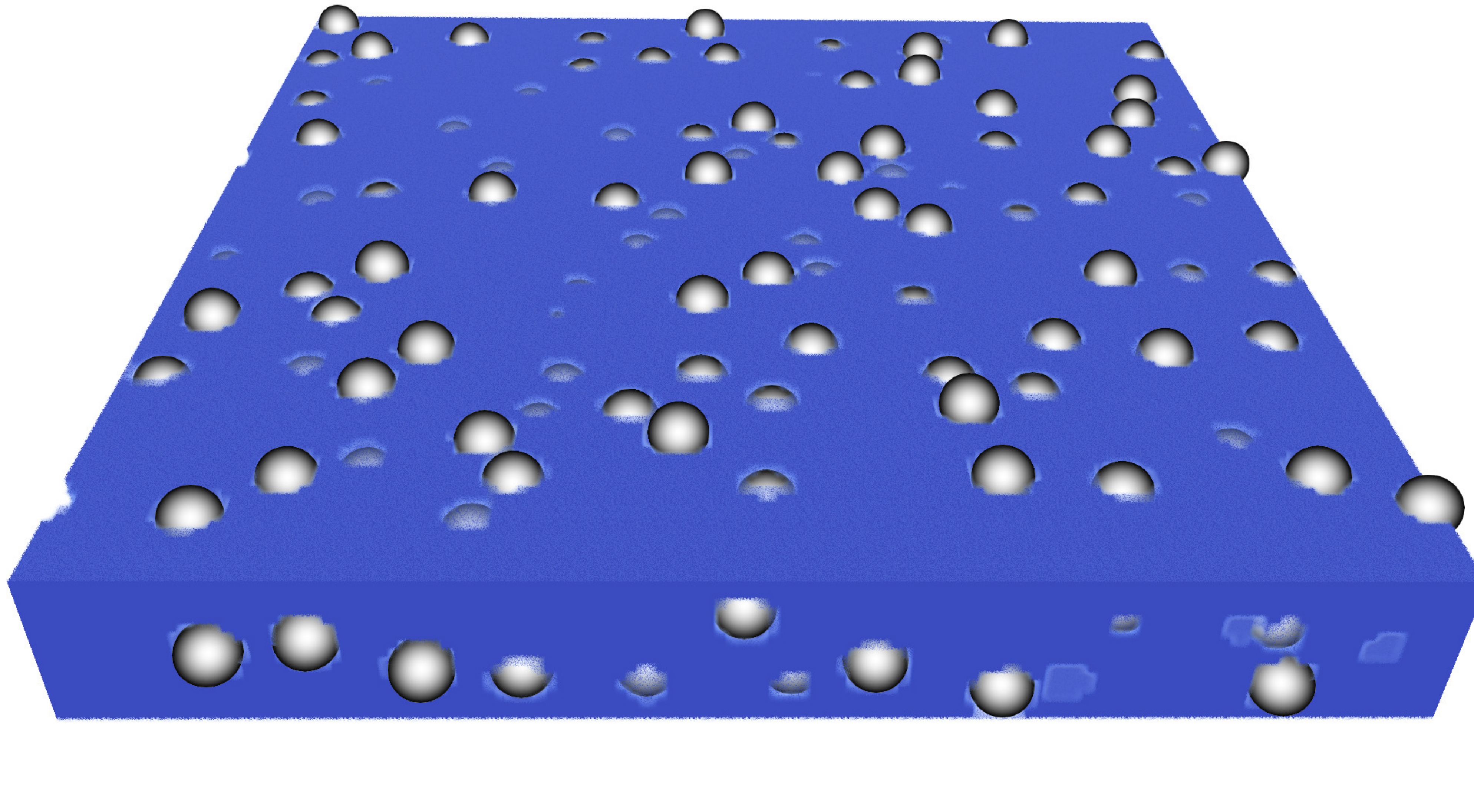}
		\subcaption{}
		\label{fig:lowtheta_1}
	\end{subfigure}
	    \begin{subfigure}{.15\textwidth}
		\includegraphics[width=0.95\textwidth]{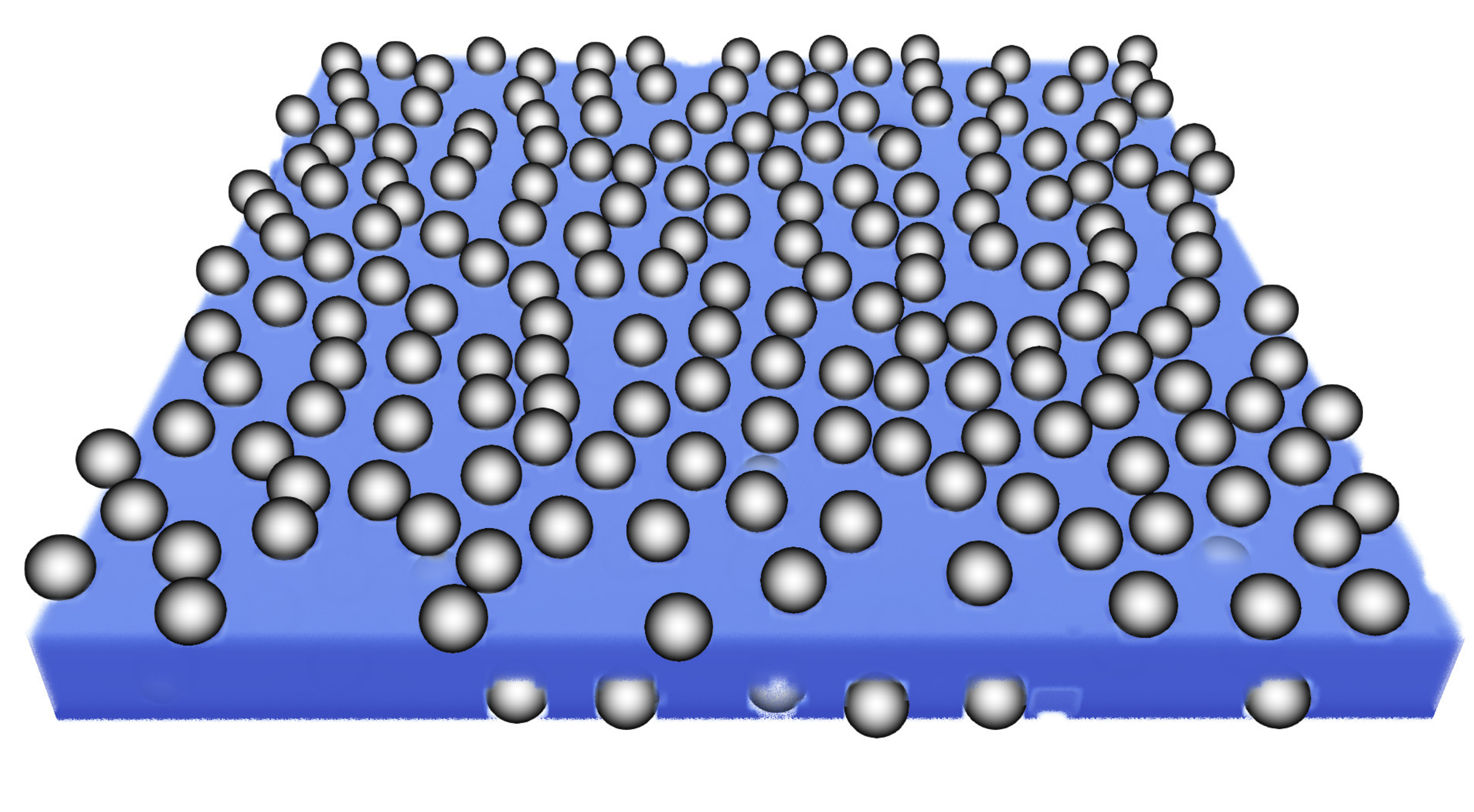}
		\subcaption{}
		\label{fig:lowtheta_2}
	\end{subfigure}
 	    \begin{subfigure}{.15\textwidth}
		\includegraphics[width=0.95\textwidth]{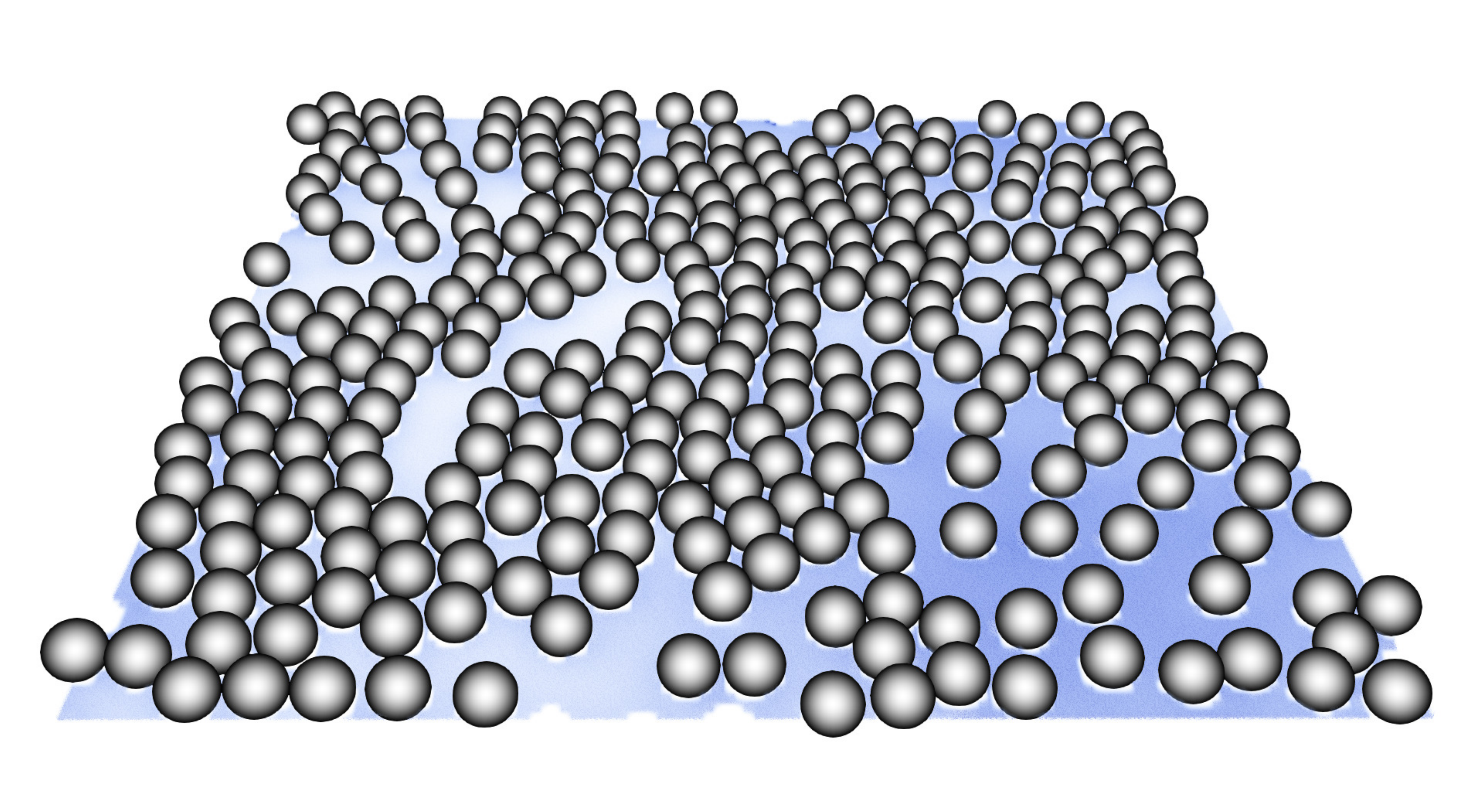}
		\subcaption{}
		\label{fig:lowtheta_3}
	\end{subfigure}
		    \begin{subfigure}{.15\textwidth}
		\includegraphics[width=0.95\textwidth]{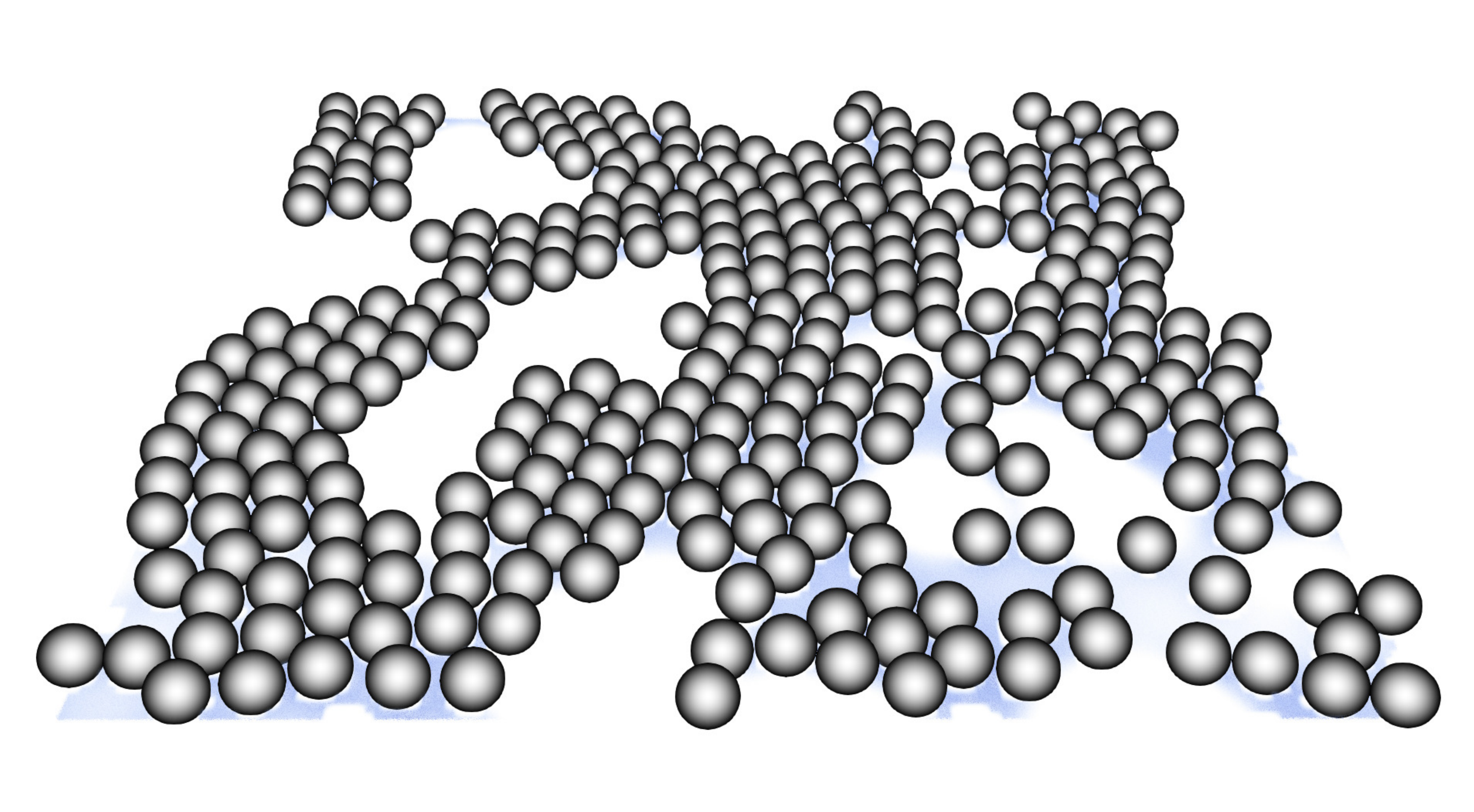}
		\subcaption{}
		\label{fig:lowtheta_4}
	\end{subfigure}
 		    \begin{subfigure}{.15\textwidth}
		\includegraphics[width=0.95\textwidth]{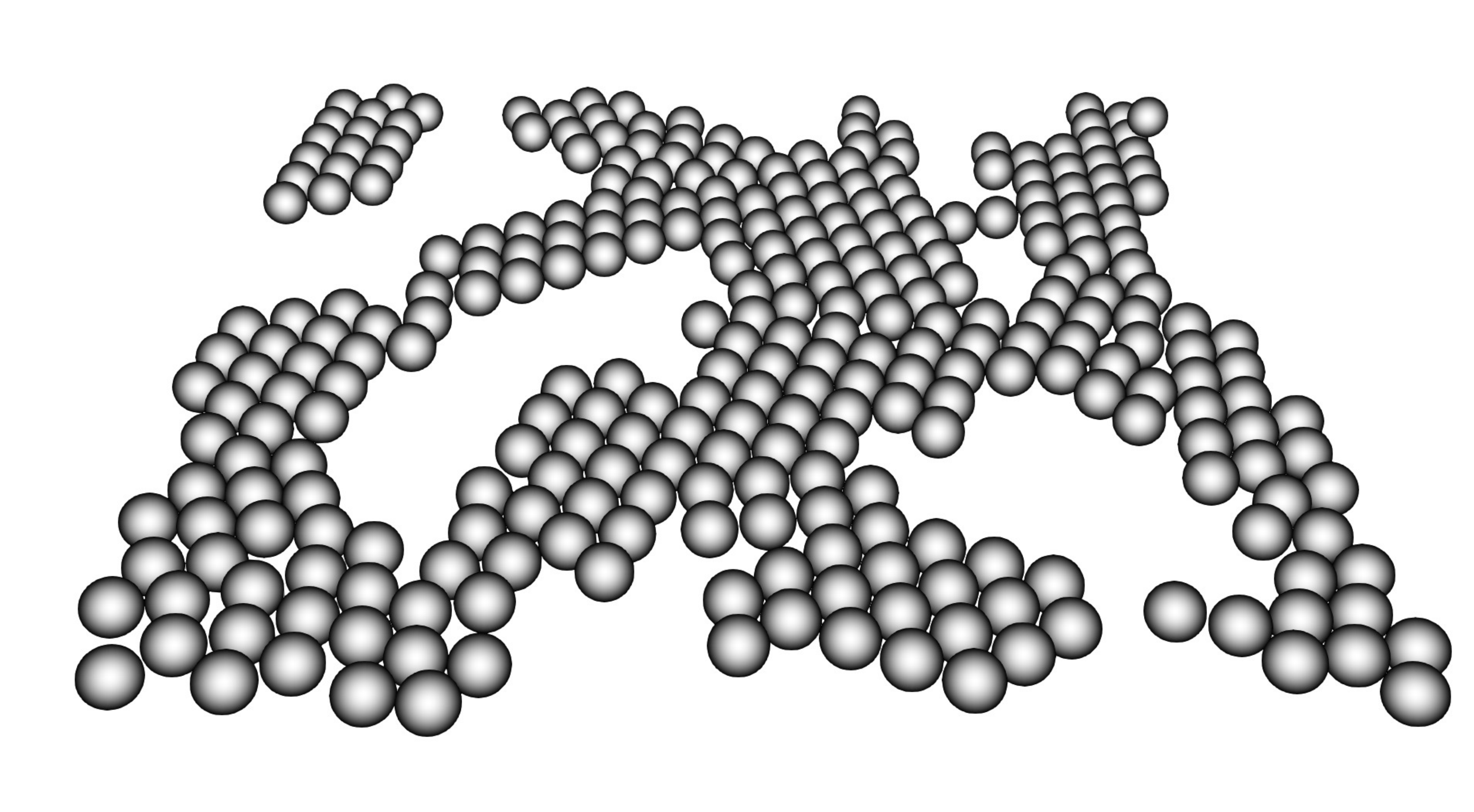}
		\subcaption{}
		\label{fig:lowtheta_5}
	\end{subfigure}

	      \begin{subfigure}{.15\textwidth}
		\includegraphics[width=0.95\textwidth]{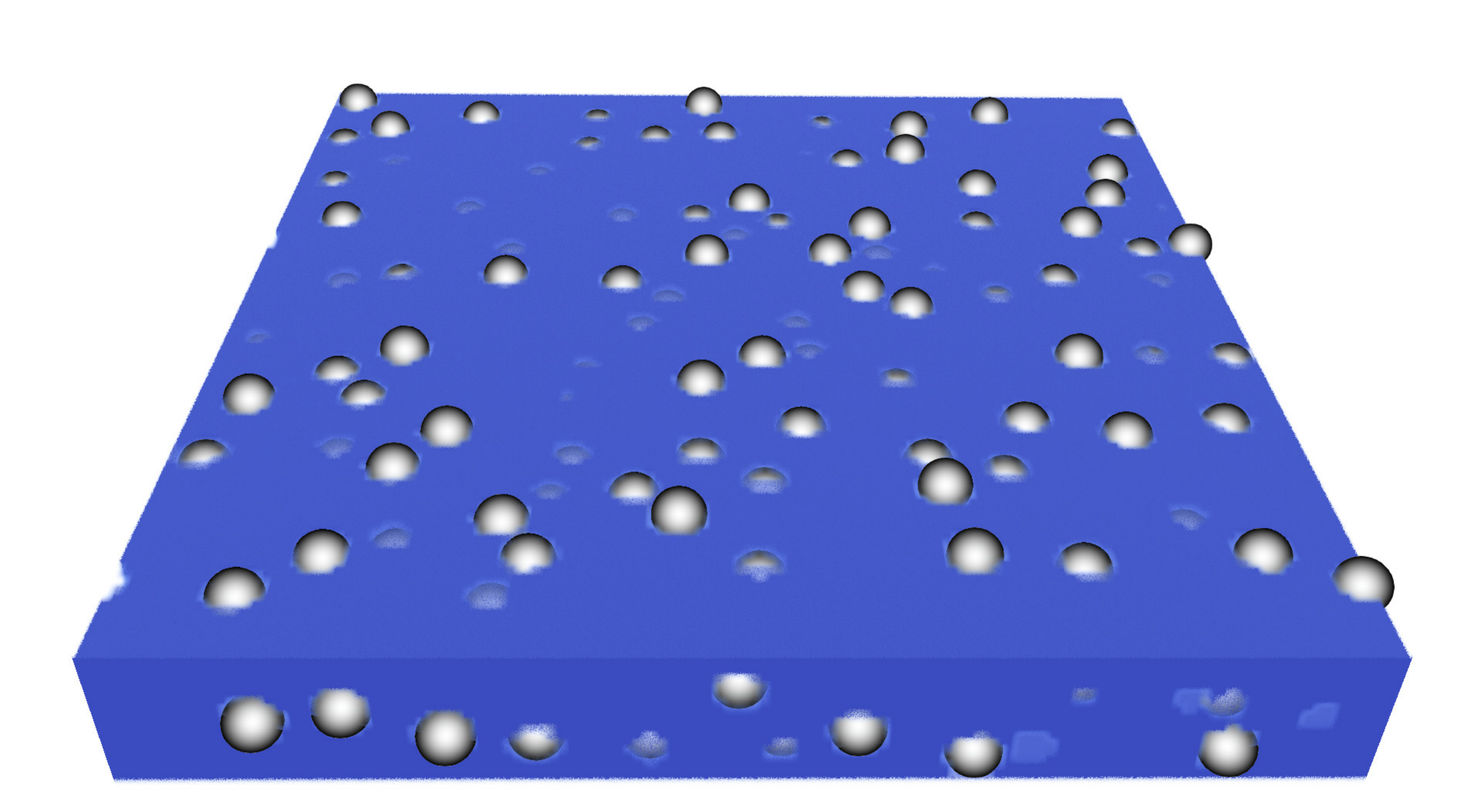}
		\subcaption{}
		\label{fig:hightheta_1}
	\end{subfigure}
	    \begin{subfigure}{.15\textwidth}
		\includegraphics[width=0.95\textwidth]{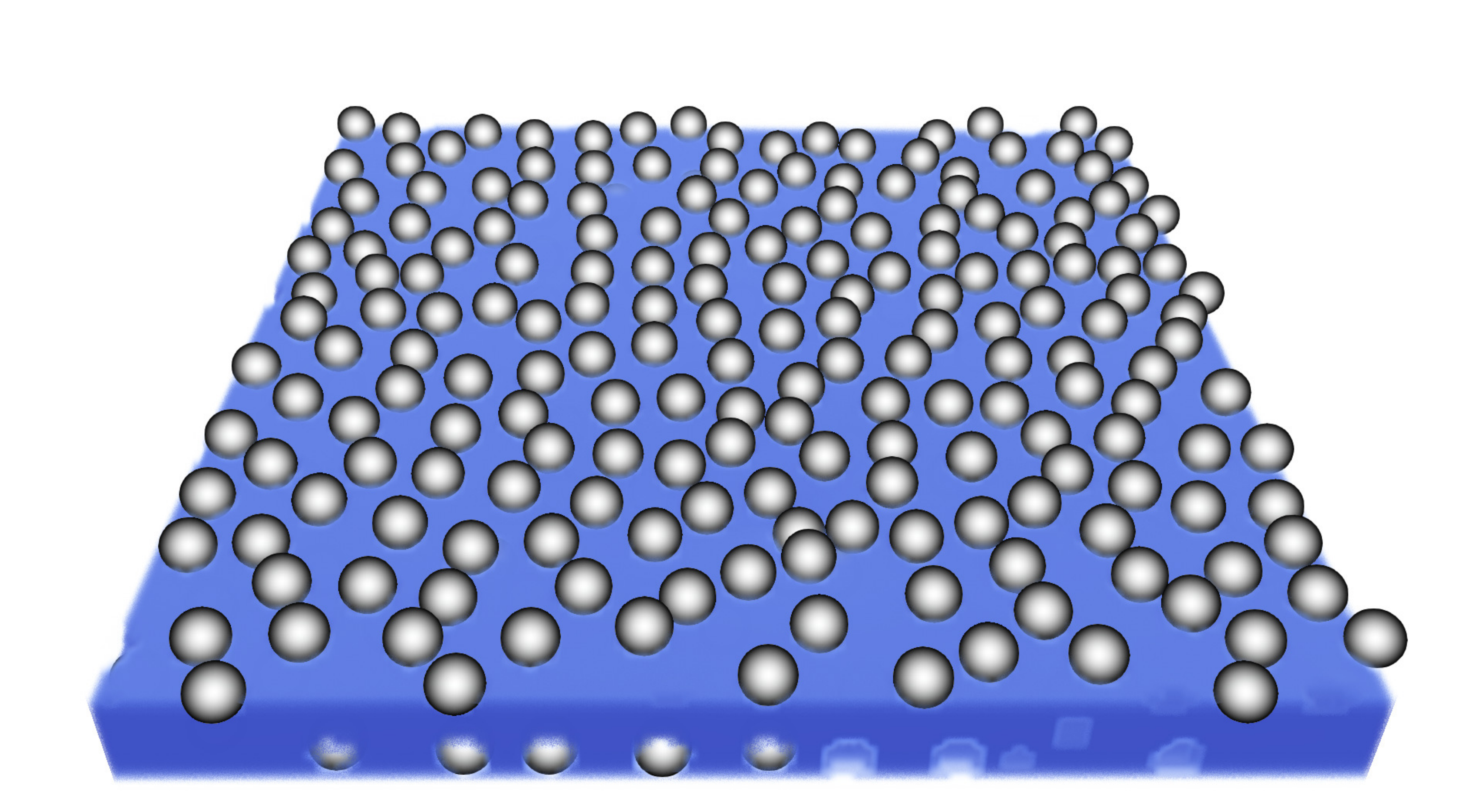}
		\subcaption{}
		\label{fig:hightheta_2}
	\end{subfigure}
 	    \begin{subfigure}{.15\textwidth}
		\includegraphics[width=0.95\textwidth]{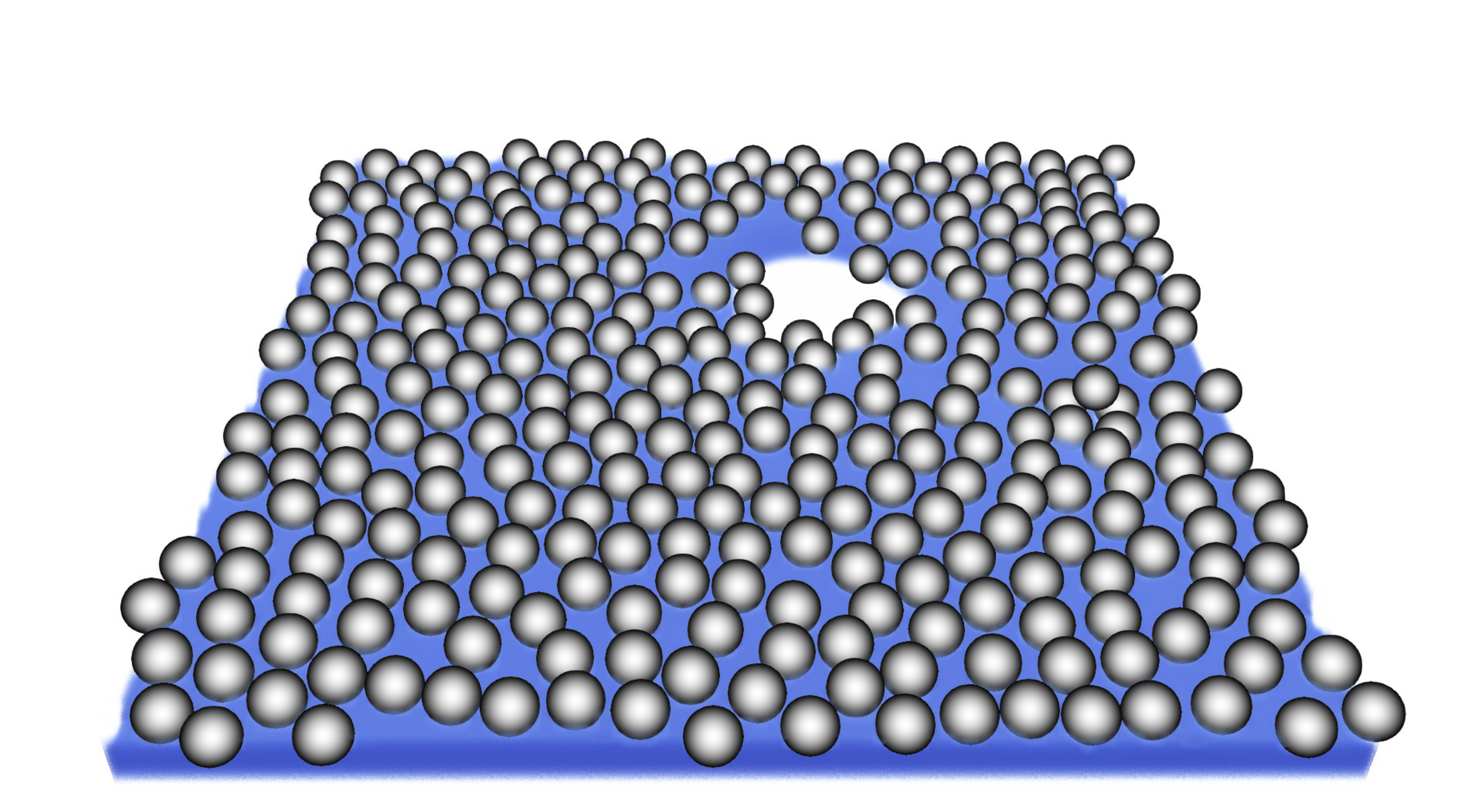}
		\subcaption{}
		\label{fig:hightheta_3}
	\end{subfigure}
		    \begin{subfigure}{.15\textwidth}
		\includegraphics[width=0.95\textwidth]{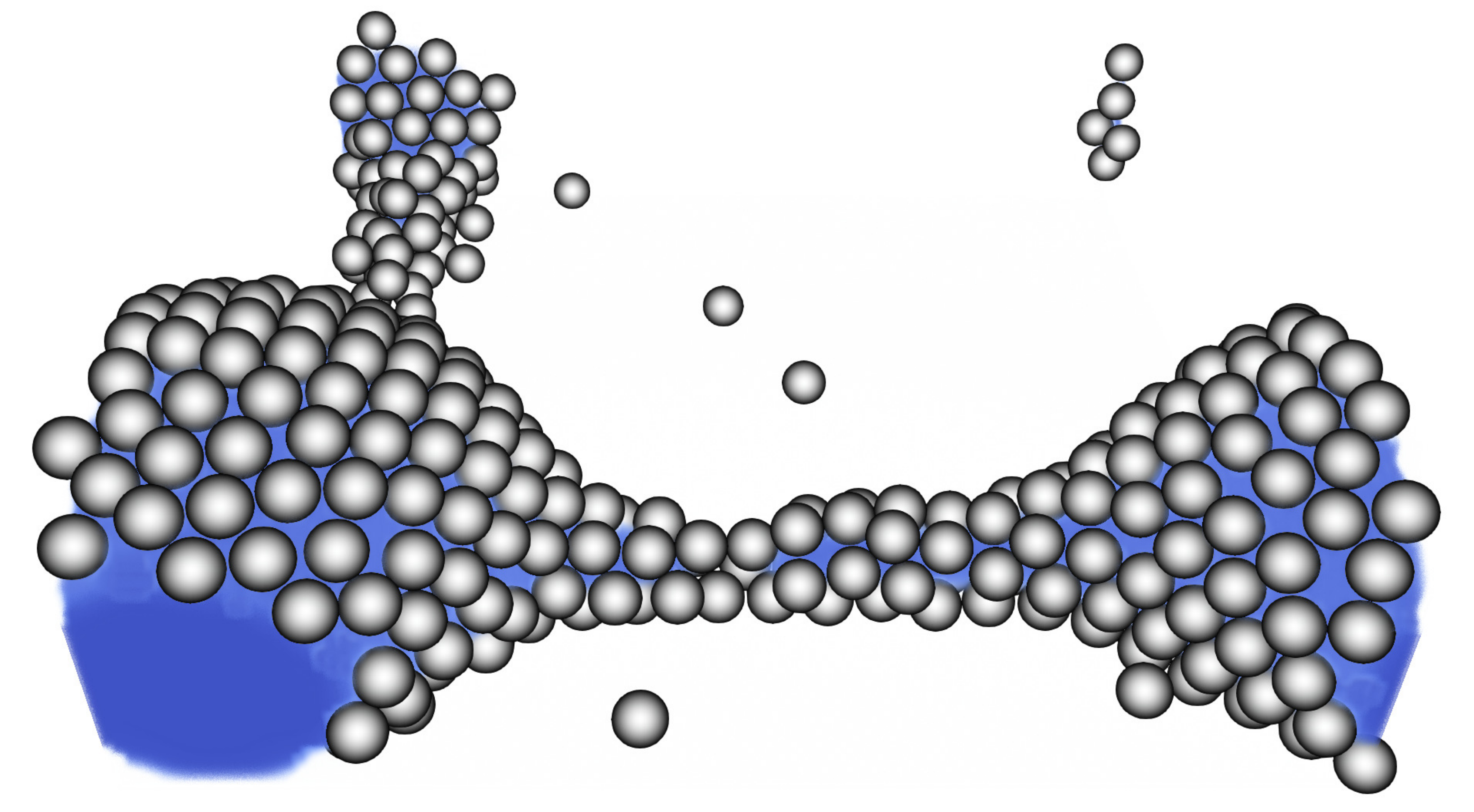}
		\subcaption{}
		\label{fig:hightheta_4}
	\end{subfigure}
 		    \begin{subfigure}{.15\textwidth}
		\includegraphics[width=0.95\textwidth]{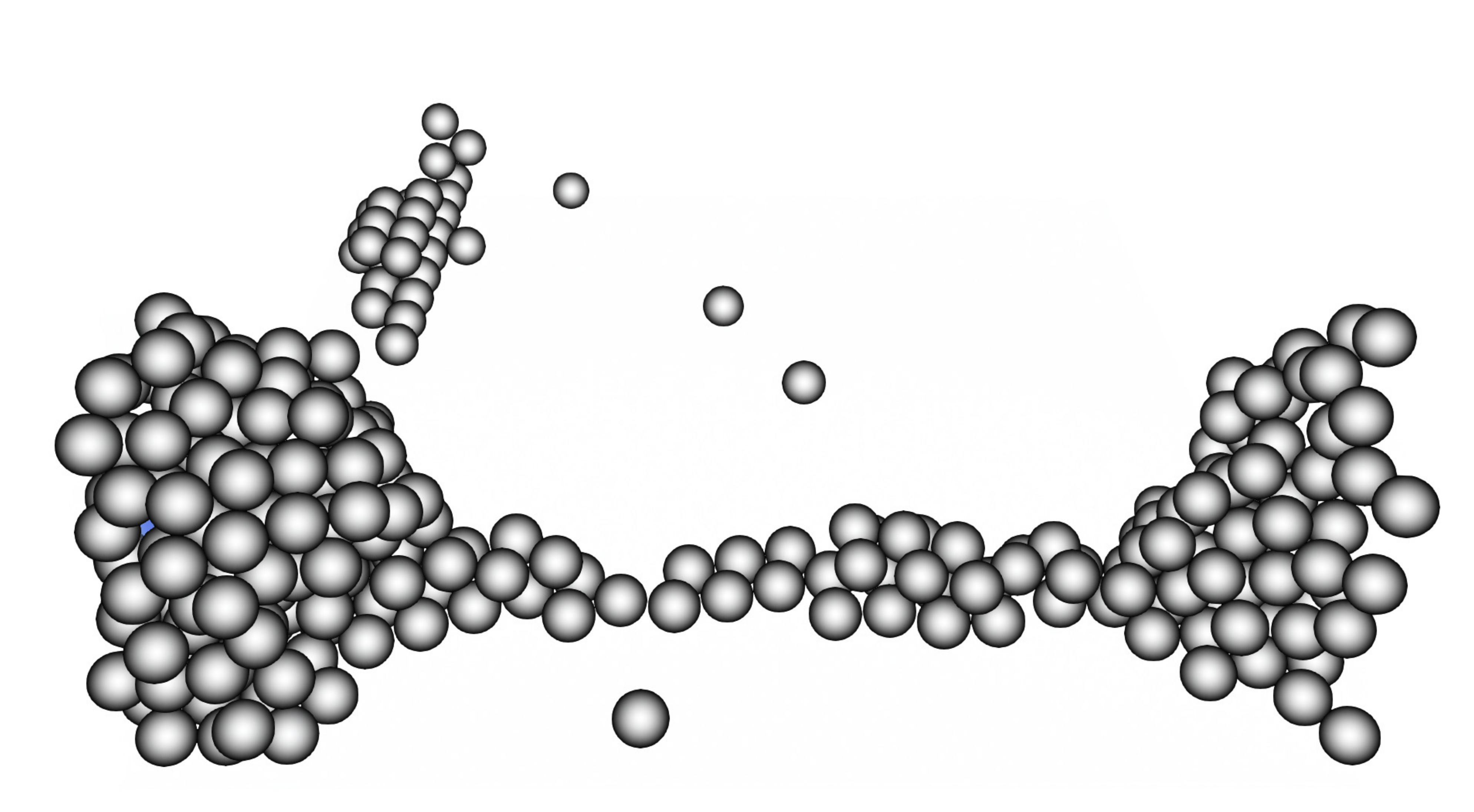}
		\subcaption{}
		\label{fig:hightheta_5}
	\end{subfigure}
     \caption{Snapshots of the drying process on a substrate with contact angles $\theta_s=30^{\circ}$ (a--e) and $\theta_s=150^{\circ} $ (f--j). The particle volume fraction is $c_v=0.15$. The fluid is represented in blue color and the particles in grey. For clarity, we omit to show the substrate. At a lower contact angle, the solvent dries and dewets resulting in capillary forces between particles, dragging the particles to form a monolayer. At a higher contact angle, droplets form after film rupture and particle clusters are left after drying.}
    \label{fig:snap-depo-theta}
\end{figure*}

\begin{figure*}[ht]
    \centering
    \captionsetup[subfigure]{justification=centering}
    \begin{subfigure}{.15\textwidth}
		\includegraphics[width=0.95\textwidth]{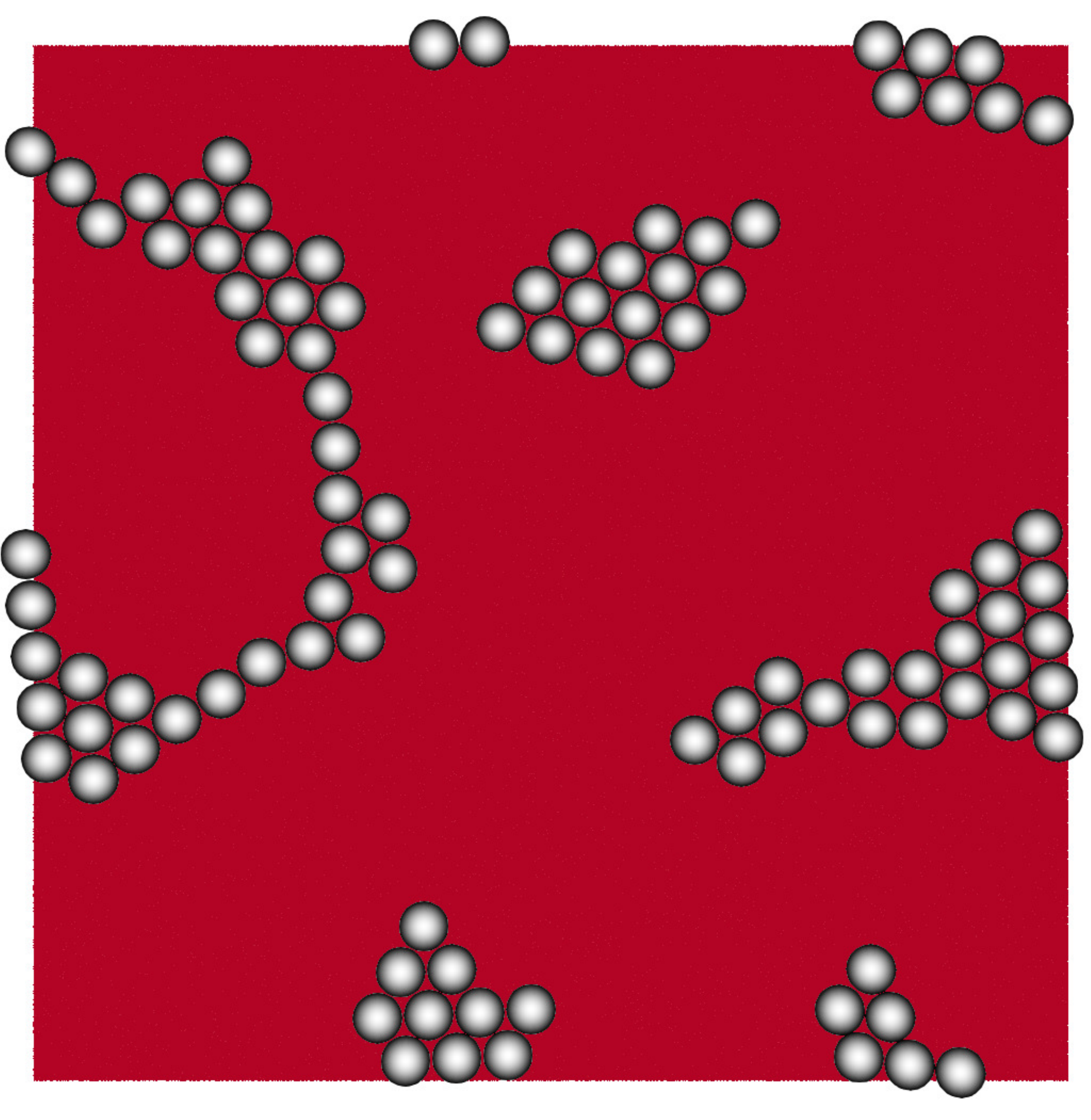}
		\subcaption{$c_v=0.04$}
		\label{fig:lowtheta_1_cv}
	\end{subfigure}
	    \begin{subfigure}{.15\textwidth}
		\includegraphics[width=0.95\textwidth]{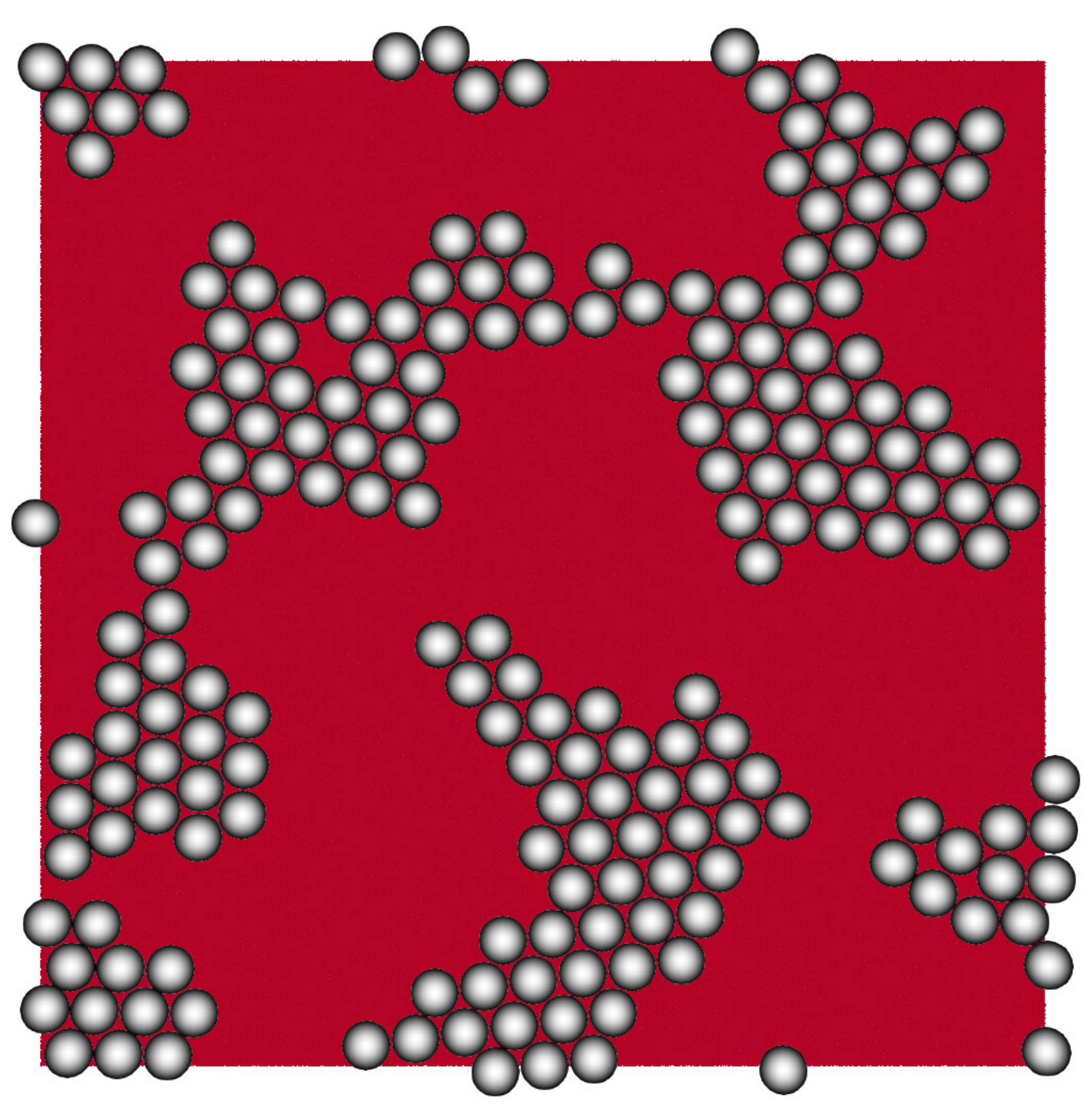}
		\subcaption{$c_v=0.08$}
		\label{fig:lowtheta_2_cv}
	\end{subfigure}
 	    \begin{subfigure}{.15\textwidth}
		\includegraphics[width=0.95\textwidth]{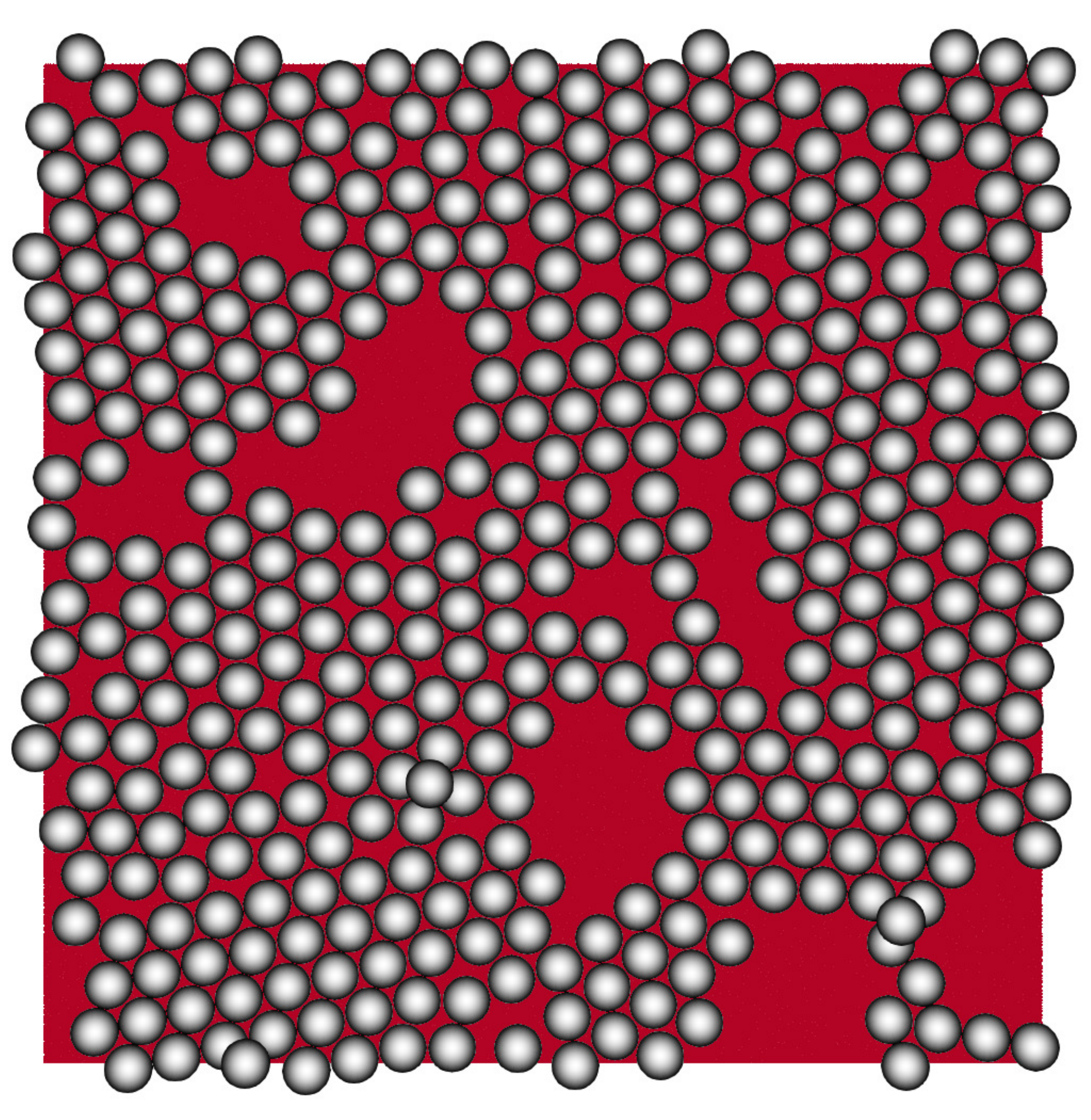}
		\subcaption{$c_v=0.15$}
		\label{fig:lowtheta_3_cv}
	\end{subfigure}
		    \begin{subfigure}{.15\textwidth}
		\includegraphics[width=0.95\textwidth]{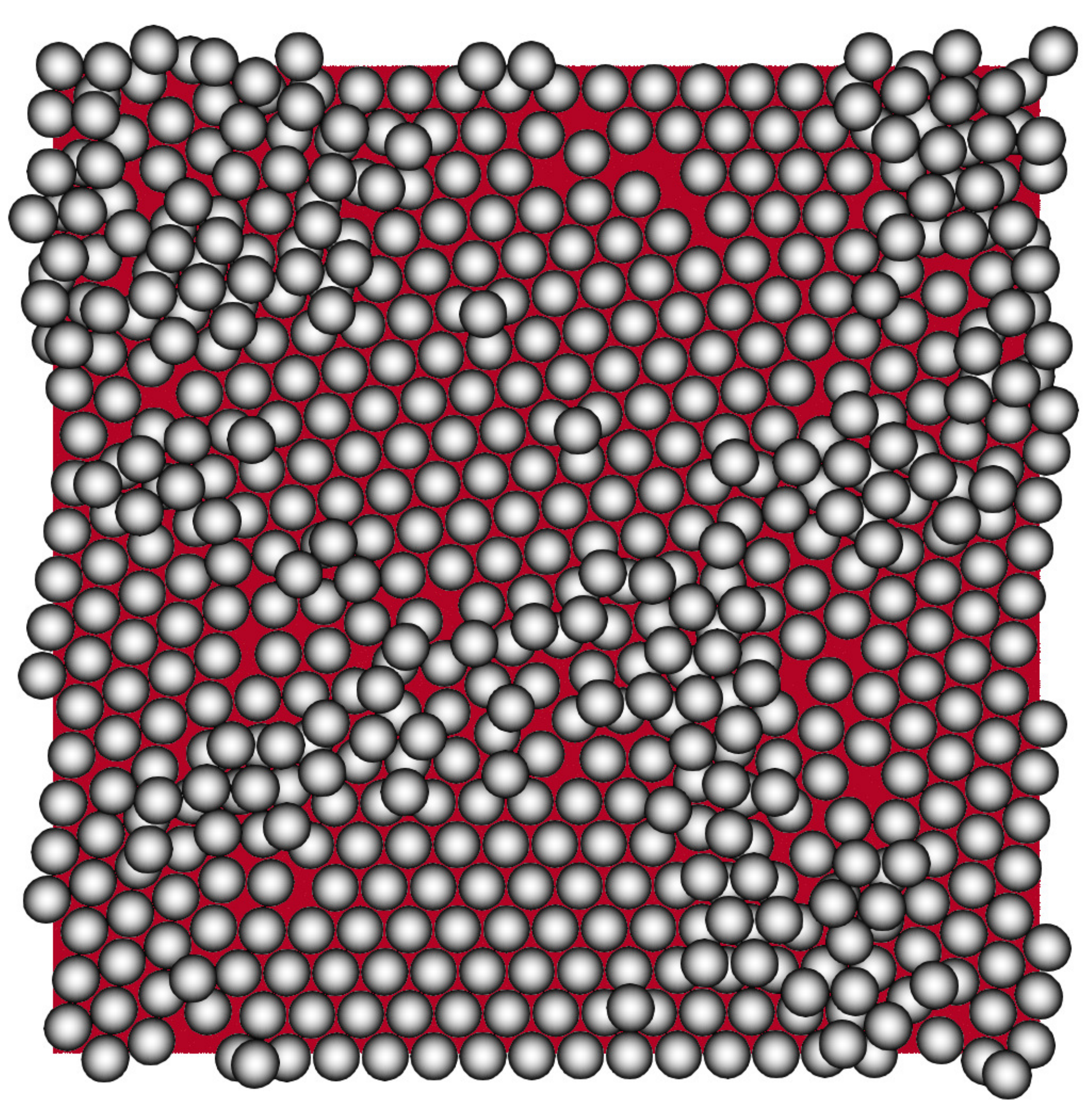}
		\subcaption{$c_v=0.23$}
		\label{fig:lowtheta_4_cv}
	\end{subfigure}

	    \begin{subfigure}{.15\textwidth}
		\includegraphics[width=0.95\textwidth]{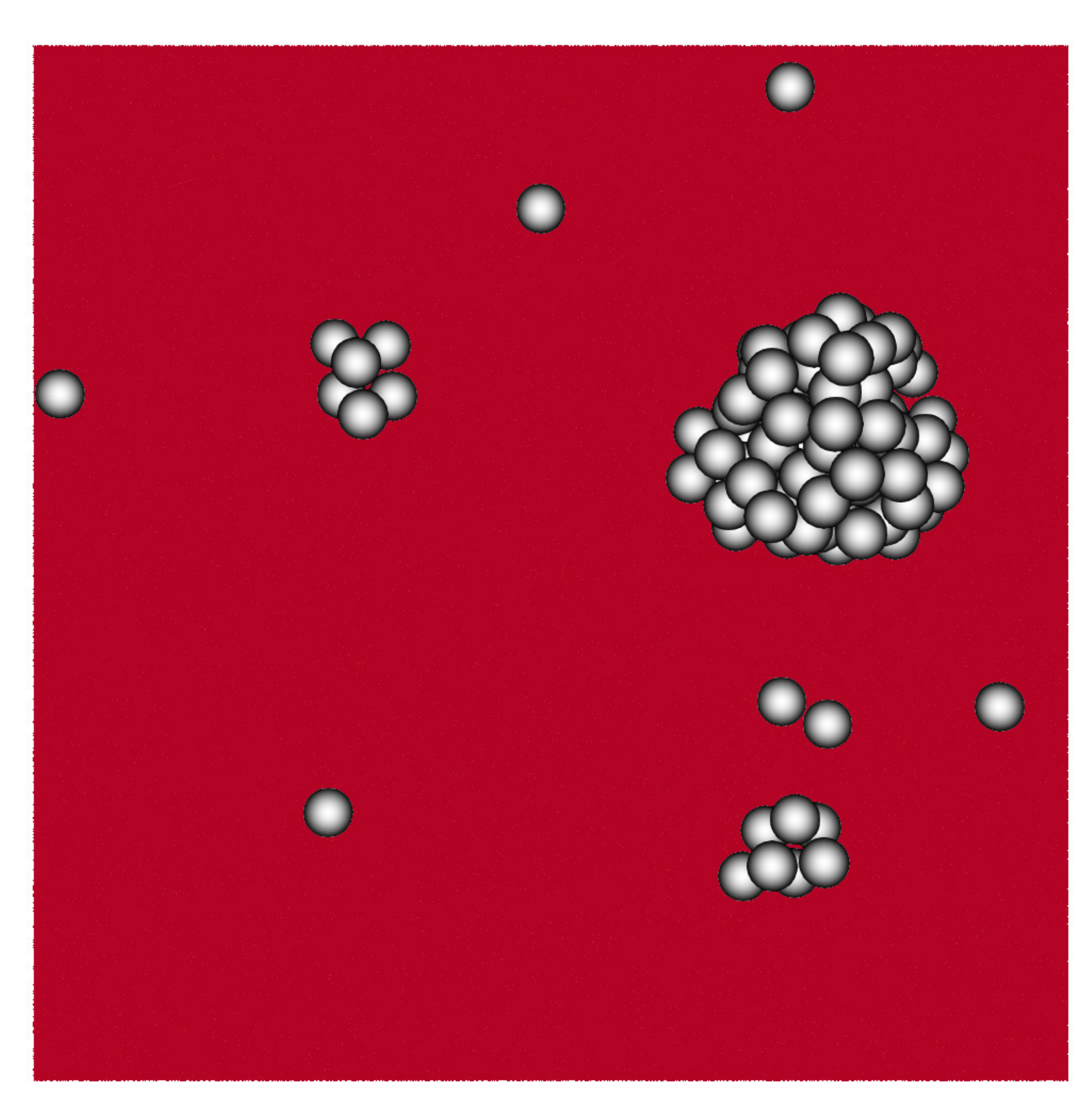}
		\subcaption{$c_v=0.04$}
		\label{fig:largetheta_1_cv}
	\end{subfigure}
	    \begin{subfigure}{.15\textwidth}
		\includegraphics[width=0.95\textwidth]{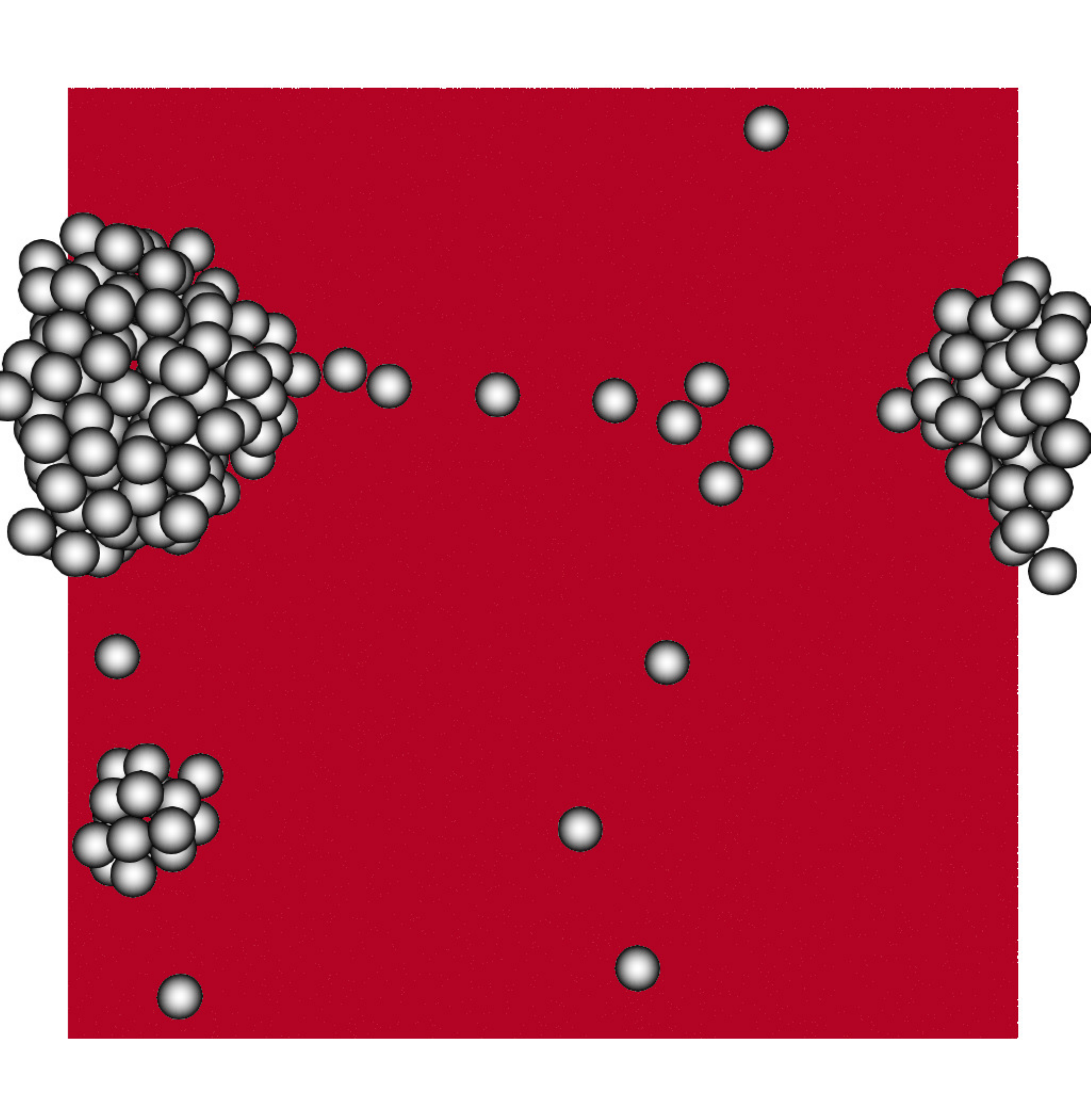}
		\subcaption{$c_v=0.08$}
		\label{fig:largetheta_2_cv}
	\end{subfigure}
 	    \begin{subfigure}{.15\textwidth}
		\includegraphics[width=0.95\textwidth]{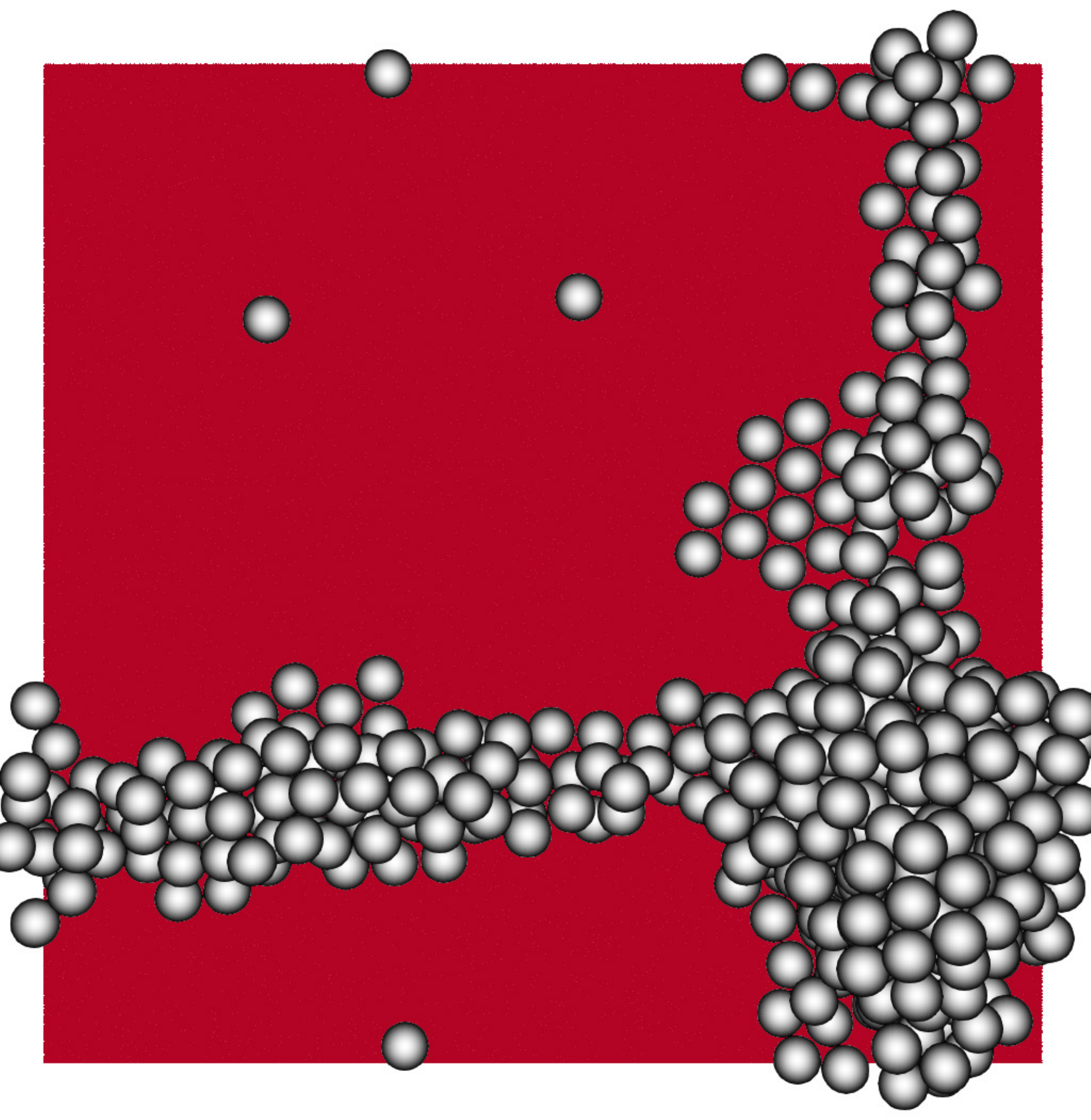}
		\subcaption{$c_v=0.15$}
		\label{fig:largetheta_3_cv}
	\end{subfigure}
	    \begin{subfigure}{.15\textwidth}
		\includegraphics[width=0.95\textwidth]{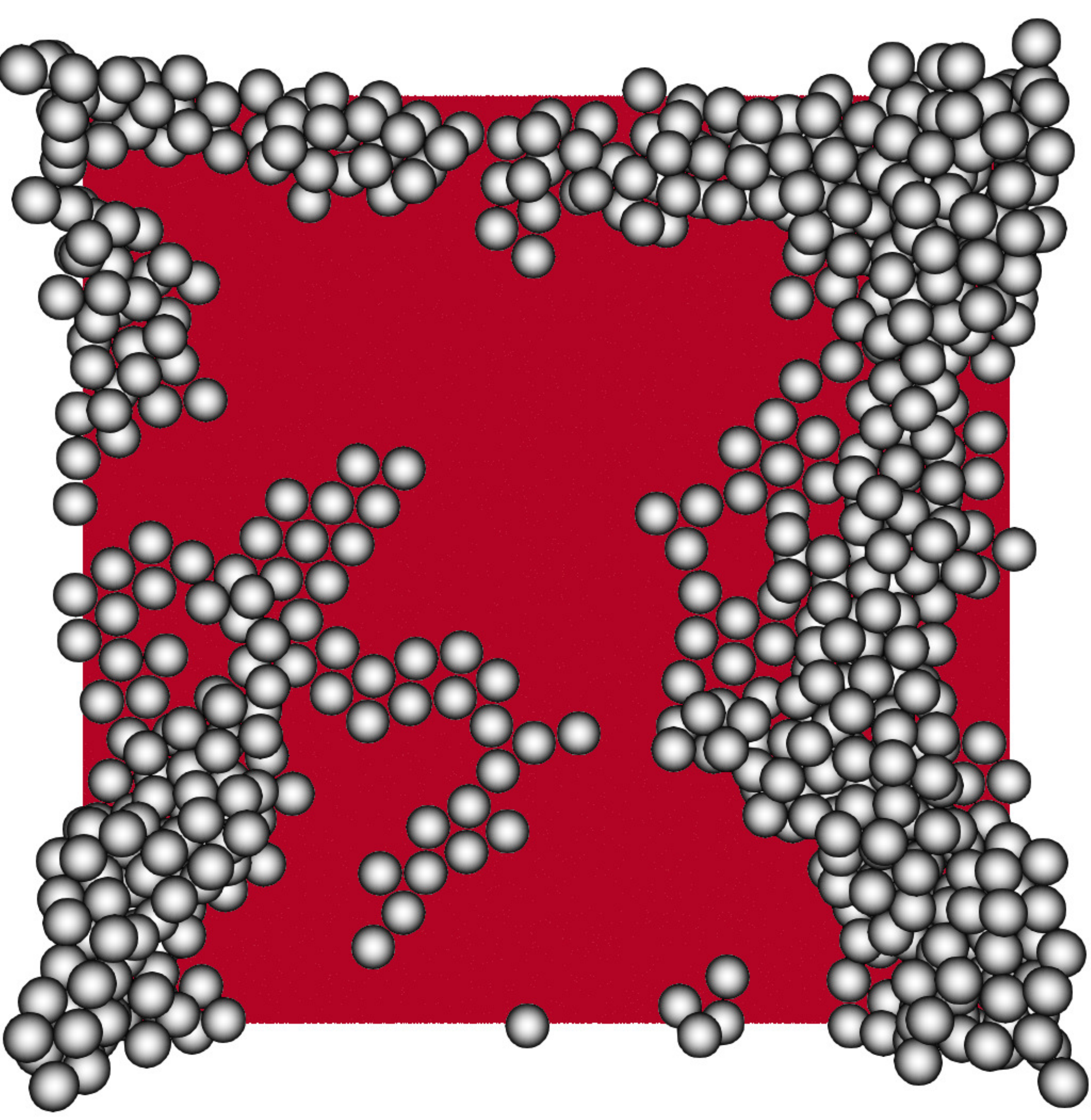}
		\subcaption{$c_v=0.23$}
		\label{fig:largetheta_4_cv}
	\end{subfigure}
    \caption{Snapshots of final deposition patterns on a substrate (shown in red) with contact angles $\theta=30^{\circ}$ (a--d) and $\theta=150^{\circ}$ (e--g) for different particle volume fractions 
    $c_v=0.04$, $c_v=0.08$, $c_v=0.15$, and $c_v=0.23$.}
    \label{fig:depo-theta_cv}
\end{figure*}

To understand the behavior of surface coverage fraction, in \figref{snap-depo-theta} we show snapshots of the drying colloidal suspension film on a substrate with contact angle  $\theta_s = 30^{\circ}$ (\figref{lowtheta_1} -- \figref{lowtheta_5}) and $\theta_s=150^{\circ}$ (\figref{hightheta_1} -- \figref{hightheta_5}), respectively. The particle volume fraction is $c_v=0.15$ and the initial height of the film is $z_0=30$.
Initially, the particles are randomly distributed in the liquid or at the interface, as shown in~\figref{lowtheta_1} and~\figref{hightheta_1}. As the drying starts, more particles get attached at the interface (\figref{lowtheta_2} and~\figref{hightheta_2}).
As the interface descends, the particles that get attached to the substrate protrude the interface and deform it. Moving forward, menisci form around the particles, giving rise to capillary forces and resulting in particle aggregation~\cite{denkov_mechanism_1992,kralchevsky_capillary_2000}. 
The aggregation of particles creates voids, ultimately leading to the rupture and dewetting of the film, as the contact line is pinned on the particle surface (\figref{lowtheta_3} and~\figref{hightheta_3}). At a lower substrate contact angle, dewetting leads to further particle aggregation (\figref{lowtheta_4}). 
Subsequently, complete evaporation of the liquid occurs, leaving a deposited monolayer on the substrate (\figref{lowtheta_5}). The particles align in a hexagonal arrangement, surrounded by areas of free particles, which is consistent with experimental observations~\cite{denkov_mechanism_1992,reculusa_synthesis_2003, perkins-howard_surface_2022,fumina_technique_2024}.
At a higher substrate contact angle, after rupture of the film, the liquid film undergoes a retraction process, rapidly forming a droplet (\figref{hightheta_4}), due to the strong repulsion between the liquid and the substrate. \revisedtext{The diffuse interface method employed here inherently accommodates topological changes in the interfacial morphology, effectively avoiding sharp curvature singularities at the rupture point.} We note that in our simulations the timescale for the film retraction to the formation of droplets is significantly shorter than the timescale of evaporation. Otherwise, the film may completely dry before forming a droplet. This behavior is consistent with that of a microscale droplet. Considering a droplet with a radius $R_d=1\mu m$, the characteristic timescale of retraction is $t_r=\sqrt{\rho_w R_{d}^{3}/\sigma_w}\approx 10^{-7}s$, which is much shorter than the characteristic evaporation timescale $t_e=\rho_w R_{d}^2/(D_w\Delta \chi) \approx 10^{-3}s$. 
Here, $D_w=2.4\times 10^{-5} m^2/s$ is the diffusion coefficient of water vapor in air, and $\Delta \chi=1.2\times 10^{-2} kg/m^3$ is the vapor concentration difference between the surface of the drop and the surroundings. 
 As the film retracts, it entrains and carries particles along, facilitating their migration onto the substrate. Subsequently, particle clusters are deposited on the substrate, as depicted in~\figref{hightheta_5}. 
The formation of droplets causes particle clustering, which likely explains the disordered arrangement of particles observed when a droplet of an aqueous suspension of monodisperse latex particles dries on hydrophobic substrates~\cite{perkins-howard_surface_2022}.

\figref{depo-theta_cv} shows the deposition pattern at different particle volume fractions $c_v=0.04$, $c_v=0.08$,
$c_v=0.15$ and $c_v=0.23$, on a substrate with a contact angle
 $\theta_s = 30^{\circ}$ (\figref{lowtheta_1_cv} -- \figref{lowtheta_4_cv}) and $\theta_s=150^{\circ}$ (\figref{largetheta_1_cv} -- \figref{largetheta_4_cv}). 
 With a lower substrate contact angle, the particles form monolayers after drying (\figref{lowtheta_1_cv} -- \figref{lowtheta_3_cv}) when the particle volume fraction is low or intermediate. With a higher volume fraction $c_v = 0.23$, the surface coverage reaches the maximal 2D packing fraction, $\phi \approx 0.77$, and additional particles can be found on top of the first deposition layer (\figref{lowtheta_4_cv}), which is also observed in experiments with higher particle volume fractions~\cite{denkov_mechanism_1992,reculusa_synthesis_2003}. 
In the case of a higher substrate contact angle, the film retracts after the rupture, forms droplets for low volume fractions, and leaves isolated particle clusters behind (\figref{largetheta_1_cv} -- \figref{largetheta_2_cv}). At higher volume fractions, finite size effects may cause the aggregates to form connected clusters (\figref{largetheta_3_cv} -- \figref{largetheta_4_cv}).

\subsection{Effect of particle wettability} 
Next, we investigate the effect of the particle wettability, characterised by the particle contact angle at the fluid-fluid interface, on the deposition process and the deposited pattern. 
The contact angle of the particles is expected to affect the pinning position of the contact line at the particle surface. 
For the following simulations, we employ again larger particles with a radius $R=6$ to suppress finite-size effects induced by the diffusive interface. 

\begin{figure}[h!]
    \centering
    \captionsetup[subfigure]{justification=centering}
    \begin{subfigure}{.45\textwidth}
    \includegraphics[width=1.0\textwidth]{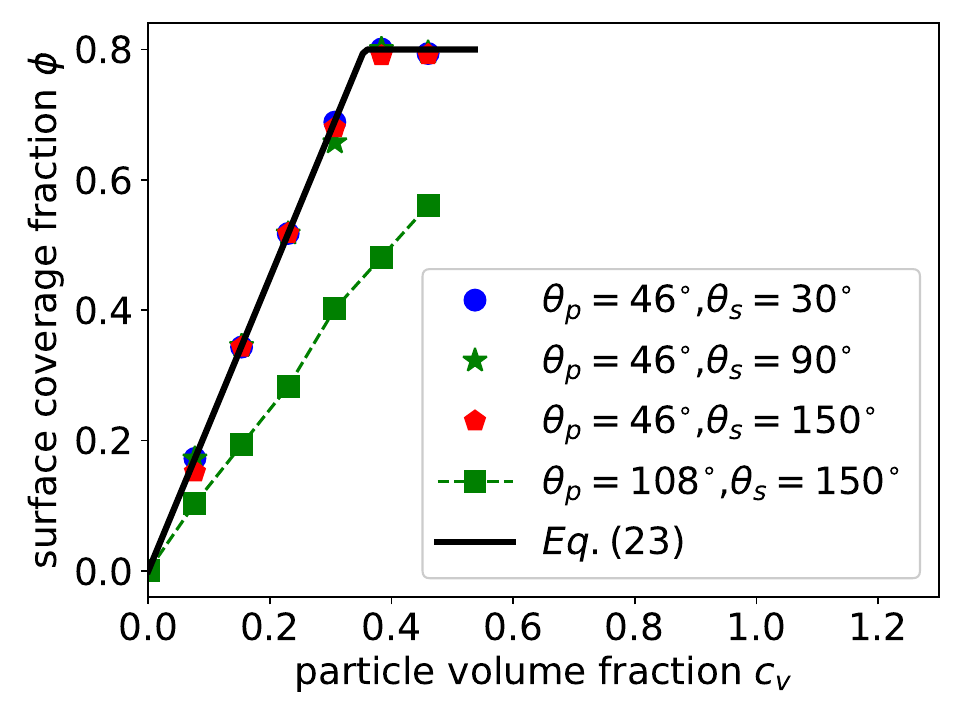}
    		\subcaption{}
		\label{fig:theta30_r6}
	\end{subfigure}
    
        \begin{subfigure}{.5\textwidth}
    \includegraphics[width=0.9\textwidth]{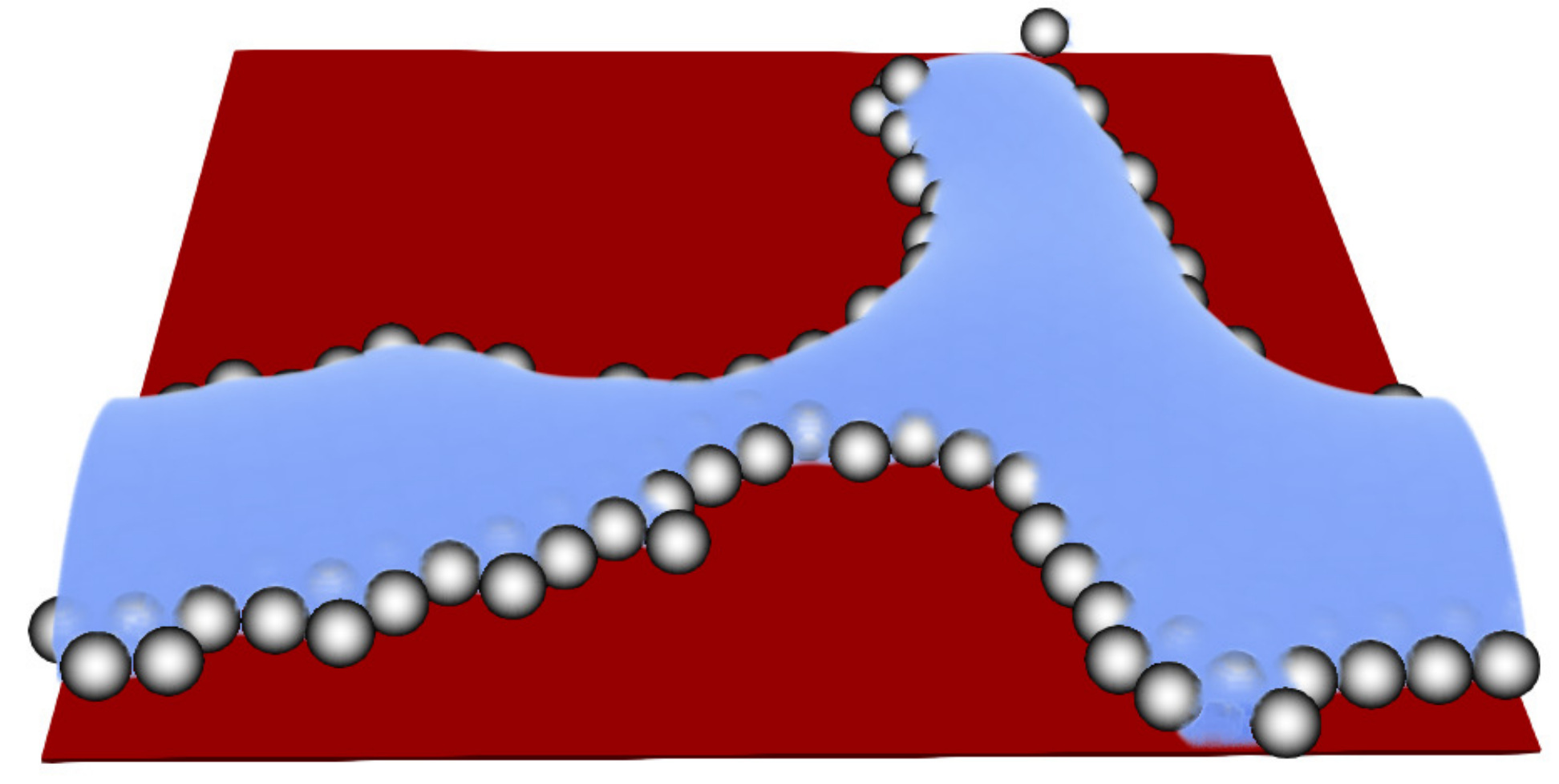}
    		\subcaption{}
		\label{fig:theta30_r6_snapshot}
	\end{subfigure}
 
    \caption{a) Surface coverage fraction as particle volume fraction $c_v$ for different substrate contact angles $\theta_s=30^{\circ}$, $\theta_s=90^{\circ}$, $\theta_s=150^{\circ}$. The radius of the spherical particles is $R=6$ and the particle contact angle is $\theta_p=46^{\circ}$. 
    b) snapshot of droplet wetting at the particle surfaces. The substrate has a contact angle $\theta_s=150^{\circ}$ and the contact angle of particles is $\theta_p=46^{\circ}$.}
    \label{fig:theta-p}
    
\end{figure}

We perform simulations of a drying colloidal suspension film on a substrate  
and compare the surface coverage fraction as a function of particle volume fraction with a lower particle contact angle $\theta_p=46^{\circ}$, for different substrate contact angles $\theta_s=30^{\circ}$ (circles), $\theta_s=90^{\circ}$ (stars) and $\theta_s=150^{\circ}$ (pentagons) (see \figref{theta30_r6}). \revisedtext{
Here, we simulate systems with higher particle volume fractions, up to
$\phi = 0.5$, compared to those shown in~\figref{theta_s}. Due to the larger particle size ($R = 6$ lattice nodes), the total number of particles was reduced by approximately a factor of eight relative to systems with the same volume fraction but smaller particles ($R = 3$ lattice nodes). As a result, the computational cost was significantly lower.}
Different from the case shown in \figref{theta_s}, where the surface coverage fraction behaves quite differently with neutral particles ($\theta_p=90^{\circ}$), here the surface coverage fraction is similar for different substrate contact angles when the particles have a lower contact angle. 
Additionally, we performed simulations using particles and a substrate with higher contact angles of $\theta_p=108^{\circ}$ and $\theta_s=150^{\circ}$, respectively. The resulting surface coverage fraction for particles with $\theta_p=108^{\circ}$ (represented by squares in \figref{theta30_r6}) is significantly lower than that obtained with particles having a lower contact angle of 
$\theta_p=46^{\circ}$ (represented by pentagons in \figref{theta30_r6}).
 
To understand this behavior, we show in~\figref{theta30_r6_snapshot} the snapshot of drying a colloidal suspension film with $\theta_p=46^{\circ}$ after rupture on a substrate with a constant angle $\theta_s=150^{\circ}$. The liquid wets at particle surfaces instead of the substrate, prohibiting the formation of a droplet and leading to a uniform, highly ordered deposit. We conclude that a lower particle contact angle eliminates the effect of substrate wettability on the deposition pattern.

\subsection{\revisedtext{Theoretical analysis}}
We propose a simple analytical analysis to predict the surface coverage fraction as a function of particle volume fraction. 
A thin film of particles with an initial thickness $z_0$ and a particle volume fraction $c_v$ is deposited on the substrate.
The total volume of particles is $V_p = Sz_0c_v$, where $S$ is the surface of the film, equal to the area of the substrate. 
Assuming that all the particles are deposited on the substrate, we expect a surface coverage area as $S_p=N_p\pi R^2$, where $N_p$ is the total number of particles. 
With $N_p=\frac{V_p}{4/3\pi R^3}$, we obtain $S_p=3Sz_0c_v/4R$.
Note that the highest surface coverage of randomly placed and equally sized spheres in a 2D arrangement is about $\phi \sim 0.77$~\cite{hinrichsen_random_1990}. 
It follows that the surface coverage fraction is given by
\begin{equation}
\phi =\frac{S_p}{S} = \begin{cases}
\frac{3z_0c_v}{4R} &\text{if $\frac{3z_0c_v}{4R} < 0.77$}\\
0.77 &\text{else}
\end{cases}
\label{eq:theory1}
\end{equation}
Based on~\eqnref{eq:theory1}, we can draw the conclusion that 
with increasing the film thickness and particle volume fraction, a smaller particle radius leads to an increased particle surface coverage fraction. 
In~\figref{theta_s} we compare the analytical prediction~\eqnref{eq:theory1} (solid lines) with our simulation results (symbols). 
For cases with lower substrate contact angles, \eqnref{eq:theory1} adequately captures the surface coverage fraction as a function of volume fraction. 
However, the model shows large deviations from the simulation results for a higher substrate contact angle. This can be attributed to particle clusters resulting from the formation of droplets.

To take into account this droplet formation, we assume that the rupture of the film is followed by a single colloidal suspension droplet being formed on the substrate with a higher contact angle. The droplet immediately reaches its equilibrium state and dries in a constant angle mode, leaving a spherical particle cluster on the substrate. We note that the contact angle of this droplet is determined by the particle contact angle if the particle contact angle is smaller than the substrate contact angle. The volume of this spherical particle cluster is
\begin{equation}
  V =  N_p  \frac{4}{3} \pi R^3 /\psi = Sz_0c_v/\psi
  \label{eq:volume1}
\end{equation}
in which $ \psi$ is the packing fraction of particles that is taken as the maximum random packing fraction of hard spheres $ \psi_{max} \approx 60\% $. 
Assuming the particle cluster has a spherical cap shape with a contact angle of $\theta$ and a footprint of $a$, we can write its volume as

\begin{equation}
    V = \frac{\pi}{6}\frac{1-\cos\theta}{\sin\theta} \left[3+\left(\frac{1-\cos\theta}{\sin\theta}\right)^2 \right] a^3 \,.
\label{eq:volume2}
\end{equation}
By combing~\eqnref{eq:volume1} and~\eqnref{eq:volume2}, we obtain the footprint of the deposit as
\begin{equation}
    a = \left( \frac{z_0Sc_v}{\frac{\pi}{6} \psi_{max} \frac{1- \cos \theta}{\sin \theta} \left[3+\left(\frac{1- \cos \theta}{\sin \theta}\right)^2 \right] } \right) ^{1/3}\,,
    \label{eq:footprint}
\end{equation}
where $\theta=\min(\theta_s,\theta_p)$.
The surface coverage fraction is
\begin{equation}
    \phi =\frac{S_p}{S} = \frac{\pi a^2}{S}\,.
     \label{eq:footprint_fraction}
\end{equation}
In~\figref{theta_s} we compare \eqnref{eq:footprint_fraction} (dashed-dotted lines) with simulation results (symbols). 
The analytical prediction agrees well with the simulations for a particle contact angle $\theta_p=90^{\circ}$ and at lower volume fractions $c_v<0.13$ on a substrate with $\theta_s=150^{\circ}$. 
The deviation at higher volume fractions is likely due to the formation of multiple droplets following film rupture in the simulations, whereas our theory assumes the formation of a single droplet.
\revisedtext{
We note that the droplet volume at rupture depends on the timing
of the rupture event. After rupture, the droplet continues to evaporate, shrinking in a spherical cap, 
until reaching a critical volume where the
particles achieve their maximum random packing fraction. \eqnref{eq:volume1} describes
this critical droplet volume with a maximal random
packing fraction. Since the final particle cluster size is determined by this critical droplet volume, the droplet volume at the rupture moment does not directly influence our analysis or the results. 
}

Our findings provide guidance for selecting appropriate solvents or substrates to form monolayers for particles with specific surface energies. The contact angle of the particles is determined by
$\cos\theta_p = \frac{\gamma_{PG}-\gamma_{PL}}{\gamma_{LG}}$ 
and the contact angle of the substrate by $\cos \theta_s = \frac{\gamma_{SG}-\gamma_{SL}}{\gamma_{LG}}$, 
where $\gamma_{ij}$ represents the surface energy between component $i$ and component $j$ and
$P, G, L, S$ denote particle, gas, liquid, and substrate, respectively.
It is preferable to choose a liquid with a lower surface energy $\gamma_{LG}$ and a substrate with a higher surface energy $\gamma_{SG}$.
Regarding the optimal volume fraction for forming a monolayer with a maximal surface coverage fraction of $0.77$ on a given substrate of area $S$, two cases are considered: i) depositing a certain amount of solution on a substrate~\cite{kaliyaraj_selva_kumar_mini-review_2020}: the solution volume is $V_d$, then the optimal volume fraction is $c_v = 1.027RS/V_d$; ii) dip-coating or blade-coating at higher coating velocities~\cite{reculusa_synthesis_2003,fumina_technique_2024}: the coated film thickness $z_0$ can be estimated using the Landau-Levich equation~\cite{LANDAU1988141,RIO2017100}, and the optimal particle volume fraction 
is then $c_v = 1.027R/z_0$ (based on~\eqnref{eq:theory1}). 

\section{Conclusion}
We numerically investigated the drying process of a colloidal suspension film on a substrate using a coupled lattice Boltzmann and discrete element method that fully resolves colloidal particles. 
This approach allows us to capture detailed information at the scale of individual particles (e.g., contact-line pinning), providing deeper insights into the deposition process as compared to existing theoretical models~\cite{pham_drying_2017,fleck_convective_2015,sobac_mathematical_2019,coombs_colloidal_2024,saiseau_skin_2025}, which typically rely on convection-diffusion equations to govern the transport of colloidal particles.

We studied the drying dynamics of a colloidal suspension film and tracked the temporal evolution of the evaporated mass. Interestingly, we found that the assembled particle monolayer at the interface does not inhibit solvent evaporation. This is because solvent transfer occurs rapidly through the particle layer, creating a saturated region above the particles that does not affect the overall evaporation flux. The evolution of film thickness during the drying of a colloidal suspension closely resembles that of a pure liquid film, consistent with our theoretical analysis. Future work should focus on the transition when the aggregation of particle multilayers begins to affect the evaporation flux~\cite{roger_controlling_2016}, which may lead to an improvement of theoretical models regarding skin formation in drying colloidal suspension droplets~\cite{daubersies_evaporation_2011, coombs_colloidal_2024, saiseau_skin_2025}.

Furthermore, we investigated the effect of substrate wettability and particle wettability on the deposition pattern. A substrate with low wettability repels the liquid, leading to the formation of droplets upon film rupture and promoting the accumulation of particles into clusters. In contrast, 
high substrate wettability facilitates better wetting and spreading of the liquid, resulting in more uniform deposition across the substrate surface. High substrate wettability proves favorable for the formation of a homogeneous monolayer.
Moreover, it is commonly believed that a hydrophilic substrate is essential for forming highly ordered monolayers in drying a film~\cite{denkov_mechanism_1992,zargartalebi_self-assembly_2022}. Surprisingly, our findings reveal that particles with high wettability can mitigate the influence of substrate wettability, as the liquid prefers to wet the particle surface instead of the substrate surface to reduce the total free energy. This facilitates the formation of highly ordered monolayers even on hydrophobic substrates. To support our simulations, we developed simple analytical models to predict the surface coverage fraction as a function of particle volume fraction, taking into account both particle and substrate wettability. The theoretical models, validated by simulation data, can be applied to predict the surface coverage fraction, potentially serving as a guide for selecting appropriate solvents or substrates to form monolayers of particles in experimental settings.

In this work, we focused on dilute suspensions of spherical particles and the formation of monolayers only.
However, our methodology can be employed directly to investigate the deposition of multiple staggered layers~\cite{singh_layerbylayer_2011,jiang_single-crystal_1999}, the effect of substrate edges~\cite{guo_guided_2001} or the impact of different particle shapes~\cite{yunker_suppression_2011,mondal_spray_2019}. Furthermore, our work can be extended to study the deposition of
 of inks involving molecules and polymers used in catalysis~\cite{liu_fabrication_2024}, batteries~\cite{lippke_simulation_2023} and biomedical applications~\cite{koppolu_development_2010}.

\begin{acknowledgements}

We acknowledge financial support from the Deutsche
Forschungsgemeinschaft (DFG, German Research Foundation) -- Project-ID 416229255 (SFB 1411) 
and Project-ID 506698391 (SPP 2196), and the German Federal Ministry of Education and Research (BMBF) -- Project H2Giga/AEM-Direkt (Grant number 03HY103HF).
We thank the Gauss Centre for Supercomputing e.V.
(\url{www.gauss-centre.eu}) for funding this project by providing computing time
through the John von Neumann Institute for Computing (NIC) on the GCS
Supercomputer JUWELS at Jülich Supercomputing Centre (JSC).
\end{acknowledgements}
\section*{Notes}
The authors declare no competing financial interest.
\section*{Data availability}
The data that support the findings of this study are openly available at
\href{http://doi.org/10.5281/zenodo.14620290}{10.5281/zenodo.14620290}. 


%

\end{document}